\documentclass[10pt,letterpaper]{article}
\usepackage{opex3}
\usepackage{amsmath}
\usepackage{amsfonts}
\usepackage{amssymb}
\usepackage{amsthm}
\usepackage{graphicx}
\usepackage{epstopdf}
\usepackage{graphicx}
\usepackage{caption}
\usepackage{subcaption}

\begin{document}

\title{A superradiant clock laser on a magic wavelength optical lattice}

\author{Thomas Maier, Sebastian Kraemer,\\Laurin Ostermann and Helmut Ritsch}

\address{Institute for Theoretical Physics, University of Innsbruck\\Technikerstrasse 25/2, A-6020 Innsbruck, Austria}

\email{laurin.ostermann@uibk.ac.at}


\begin{abstract} 
An ideal superradiant laser on an optical clock transition of noninteracting cold atoms is predicted to exhibit an extreme frequency stability and accuracy far below mHz-linewidth. In any concrete setup sufficiently many atoms have to be confined and pumped within a finite cavity mode volume. Using a magic wavelength lattice minimizes light shifts and allows for almost uniform coupling to the cavity mode. Nevertheless, the atoms are subject to dipole-dipole interaction and collective spontaneous decay which compromises the ultimate frequency stability. In the high density limit the Dicke superradiant linewidth enhancement will broaden the laser line and nearest neighbor couplings will induce shifts and fluctuations of the laser frequency. We estimate the magnitude and scaling of these effects by direct numerical simulations of few atom systems for different geometries and densities. For Strontium in a regularly filled magic wavelength configuration atomic interactions induce small laser frequency shifts only and collective spontaneous emission weakly broadens the laser. These interactions generally enhance the laser sensitivity to cavity length fluctuations but for optimally chosen operating conditions can lead to an improved synchronization of the atomic dipoles.
 \end{abstract}


\bibliographystyle{osajnl}

\section{Introduction}

An essential and characteristic property of laser light, observed since its first generation, is its extraordinary coherence and frequency stability well below the width of the optical resonator used. Far above threshold the linewidth is limited by technical noise of the gain medium and the mirrors only. Continuous technological advances have brought this limit down to an incredible stability below the Hz-level~\cite{kessler2012sub}, which competes against the $Q$ and the linewidth of long lived atomic clock states. At this point, further technological improvements seem extremely challenging. Therefore, it has been suggested recently~\cite{meiser2009prospects} and to some extent demonstrated experimentally~\cite{bohnet2012steady,bohnet2013active} that an atomic clock transition could be used as a narrow band gain medium to run a laser. Due to the very feeble individual dipole moments of the atoms such a device can only be operated in the strong collective coupling regime, where superradiant emission into the field mode provides for the necessary gain~\cite{vuletic2012atomic}. In this domain of operation a huge collective dipole constituted by a large number of atoms, which are synchronized via their common coupling to the cavity field~\cite{xu2013simulating,henschel2010cavity}, will build up.

The general idea of superradiant lasers and their properties have been discussed already two decades ago~\cite{haake1993superradiant,horak1995quantum}, where a unique frequency stability scaling with the inverse square of the atom number $N$ and squeezed output light was predicted. Their superb accuracy in the regime of a cavity linewidth much larger than the atomic linewidth were highlighted just recently~\cite{bohnet2012steady}. Most importantly, in this case the laser becomes very insensitive to technical noise in the resonator and its properties are dominated by the intrinsic stability of the collective atomic dipole. Under favorable conditions, with only a few photons and millions of atoms present, a natural width of the system several orders of magnitude below the $1$ Hz-level could be envisaged.

A central, yet open technical problem here is the implementation of a uniform collective coupling of the atoms to the field mode as well as the optical pumping in the atomic system without a considerable perturbation of the lasing levels, which in this case include the atomic ground state. Thus, a very careful choice of operating parameters is required. Here, we study another intrinsic source of perturbation in this sensitive system, namely direct dipole-dipole interaction between the laser active atoms, as they are densely confined within the optical resonator. Similar to atom-atom collisions in Ramsey experiments~\cite{swallows2011suppression}, dipole-dipole couplings tend to induce phase noise and decoherence of the collective atomic dipole. In particular in lattice setups at low filling, where collisions are strongly suppressed, this should constitute the most prominent source of noise for such a laser.

The basic phenomenon of superradiance was theoretically studied in detail, e.g. by Haroche and coworkers~\cite{intro-haroche}, about 50 years ago using a variety of analytical approximation methods~\cite{bonifacio1971quantum, intro-eberly}. As an important effect one finds that the decay rate of low energy collective excitations grows linearly with the particle number $N$~\cite{zoubi2012collective}. For multiply excited states the effect is increased further and the collective decay of a strongly inverted ensemble exhibits a delayed intensity maximum largely proportional to $N^2$ as a significant deviation from the exponential decay of individual atoms~\cite{macgillivray1976theory,intro-haroche,Ostermann2012Cascaded}. The phenomenon has been observed in a large number of experiments in gases and solids~\cite{intro-skribanowitz, macgillivray1976theory} and more recently also for ultracold quantum gases~\cite{inouye1999superradiant,moore1999theory}. 

As in every laser setup, we naturally have to deal with inverted ensembles. Hence, we can expect that superradiant effects will play an important role and the assumption of individual atomic decay at the independent free space single atom rate will loose its validity. Let us emphasize that the collective symmetric coupling of the atoms to the cavity mode does not require the atoms to occupy a small volume of the order of a cubic wavelength, but simply calls for an almost equal cavity coupling constant for all atoms. Dicke superradiant spontaneous decay, on the other hand, is maximal for closely spaced emitters, but still plays a decisive role in more extended geometries and in particular for regularly ordered ensembles. While this free space superradiant interaction and decay was incorporated intrinsically in the early works on superradiant lasing~\cite{haake1993superradiant}, it was neglected in the more recent considerations on superradiant lasing on ultra narrow atomic transitions~\cite{meiser2009prospects}.

In the present paper we investigate the full model for the collective decay process in a superradiant laser configuration. While the underlying Hamiltonian and dynamical master equations for the coupled atom-field dynamics are well established, exact treatments of the full decay problem for more than a few particles is hardly possible apart from some special cases. Numerical simulations can be performed for somewhat higher atom numbers in the fully collective limit. However, for more particles in a small but finite volume, collective and individual decay are present and the equations immediately become very cumbersome, as the number of occupied states within the total physical Hilbert space (growing as $2^N$) gets prohibitively large. Interesting results can still be obtained for special finite configurations, which should exhibit the qualitative consequences of dipole-dipole coupling quite well.  Besides demonstrating the underlying basic physical mechanisms, our study aims at direct implications for the laser linewidth of a magic wavelength lattice laser in the superradiant regime~\cite{takamoto2005optical,casperson1973spectral,bohnet2012steady,henschel2010cavity}.

\section{Model}

We consider $N$ identical two-level atoms held in a regular spaced configuration, e.g. in a far detuned optical trap, each of them symmetrically coupled to a single mode of a high $Q$ optical resonator. Due to the inherent exposure of the atoms to the vacuum bath the ensemble is affected by coherent dipole-dipole energy exchange processes and also by collective spontaneous emission~\cite{Lehmberg1970}. Further, we employ a transverse incoherent pump, which allows us to use the atoms as an active medium, as well as another dissipative process, the cavity loss. Upon Born, rotating wave and Markov approximation we end up with a standard Lindblad type master equation. Explicitly the time-dependence of the $N$-atom density matrix  is governed by ($\hbar = 1$)
\begin{equation}
\frac{\partial \rho}{\partial t} = i \left[ \rho, H \right] + \mathcal{L}_{\text{cd}} \left[ \rho \right] + \mathcal{L}_{\text{pump}} \left[ \rho \right] + \mathcal{L}_{\text{cav}} \left[ \rho \right] = \mathcal{L} \left[ \rho \right],
\label{mastereq}
\end{equation}
with the Hamiltonian
\begin{equation}
H = \frac{\omega_0}{2} \, \sum_i \sigma_i^z + \sum_{i \not = j} \Omega_{ij} \, \sigma_i^+ \sigma_j^- + \omega_c \, a^\dagger a + H_{\text{int}},
\label{H_dipole}
\end{equation}
where $\sigma_i^+$ and $\sigma_i^-$ are the raising and lowering operators for the atomic dipole of the $i$-th atom with the transition energy $\omega_0$, the operators $a^\dagger$ and $a$ correspond to the creation and annihilation of a photon with the frequency $\omega_c$ in the cavity mode , $\Omega_{ij}$ denotes the resonant dipole-dipole energy transfer between the atoms $i$ and $j$, and
\begin{equation}
H_{\text{int}} = g \sum_i \left( a \sigma_i^+ + a^\dagger \sigma_i^- \right)
\end{equation}
represents the Jaynes-Cummings type interaction between the individual atomic transition dipoles and the cavity mode with $g$ being the coupling that emerges if a constant mode function is assumed. This approximation is justified in the situation where the atomic ensemble is aligned transversely to the propagation direction of the cavity mode or its dimensions are much smaller than the length of the resonator.

\begin{figure}
\includegraphics[width=\textwidth]{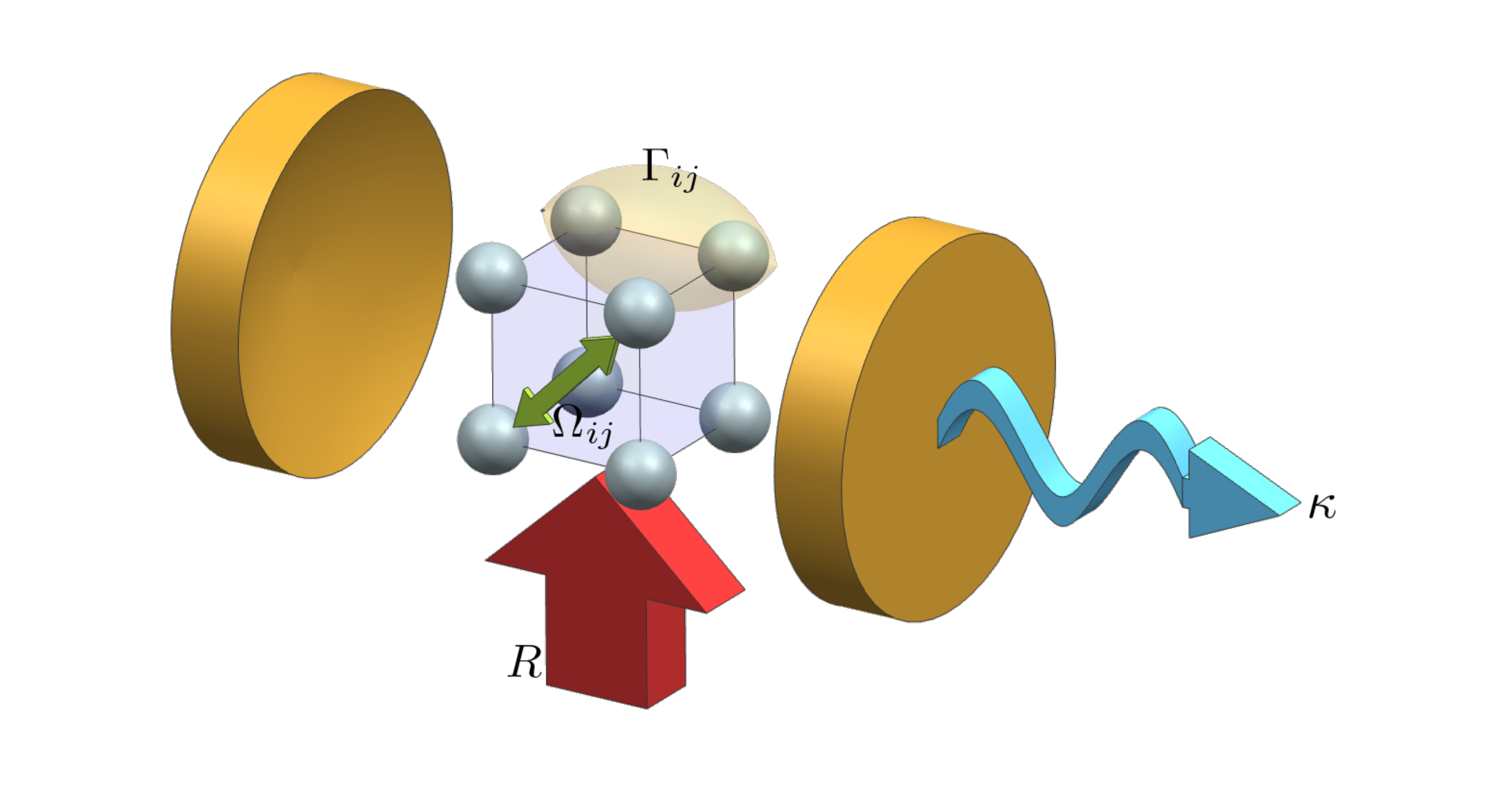}
\caption{Schematics of a lattice laser setup. A transversely pumped (pumping rate $R$) finite atomic ensemble with dipole-dipole couplings $\Omega_{ij}$ and collective spontaneous emission $\Gamma_{ij}$ inside an optical resonator with a loss rate of $\kappa$}
\end{figure}

The collective atomic damping is accounted for by the Liouvillian
\begin{equation}
\mathcal{L}_{\text{cd}} \left[ \rho \right] = \frac{1}{2}  \sum_{i, \, j} \Gamma_{ij} \left( 2 \sigma_i^- \rho \sigma_j^+ - \sigma_i^+ \sigma_j^- \rho - \rho \sigma_i^+ \sigma_j^- \right)
\label{L_decay}
\end{equation}
with generalized spontaneous emission rates $\Gamma_{ij}$ arising from the coupling of the atomic transition dipoles through the vacuum field~\cite{Ficek1987}. The incoherent transverse broadband pumping, which in our model acts on each atom in the same way, leads to
\begin{equation}
\mathcal{L}_{\text{pump}} \left[ \rho \right] = \frac{R}{2} \sum_{i} \left( 2 \sigma_i^+ \rho \sigma^- - \sigma^-_i \sigma^+_i \rho - \rho  \sigma_i^- \sigma_j^+ \right) 
\end{equation}
with $R$ quantifying the pumping rate and cavity loss with the rate $\kappa$ is described by
\begin{equation}
\mathcal{L}_{\text{cav}} \left[  \rho \right] = \kappa \left(2 a \rho a^\dagger  - a^\dagger a \rho - \rho a^\dagger a \right).
\end{equation}

Observe that the collective coupling and decay matrices $\left[ \Omega_{ij} \right]$ and $\left[ \Gamma_{ij} \right]$ possess non-diagonal elements, which have to be calculated as a function of the system's geometry~\cite{Ostermann2012Cascaded}. In many other cases, due to the finite correlation length of vacuum fluctuations, these nondiagonal parts can be safely neglected. Explicitly, for identical atoms we have~\cite{intro-ficek}
\begin{equation}
\Gamma_{ij} =  \frac{3 \Gamma}{2} F \left( k _0 r_{ij} \right) \qquad \Omega_{ij} =  \frac{3 \Gamma}{4} G \left( k_0 r_{ij} \right)
\end{equation}
with $\Gamma$ the single atom linewidth, $k_0 = \omega_0/c = 2 \pi/\lambda_0$ and
\begin{equation} \begin{aligned}
F \left( \xi \right) &= \left( 1- \cos^2 \theta \right) \frac{\sin \xi }{\xi} + \left( 1- 3 \cos^2 \theta \right) \left( \frac{\cos \xi}{\xi^2 }- \frac{\sin \xi}{\xi^3} \right),\\
G \left( \xi \right) &= - \left( 1- \cos^2 \theta \right) \frac{\cos \xi }{\xi} + \left( 1- 3 \cos^2 \theta \right) \left( \frac{\sin \xi}{\xi^2}+\frac{\cos \xi}{\xi^3} \right),
\end{aligned} \end{equation}
where $\xi = k_0 r_{ij}$. Here, $r_{ij}$ denotes the relative distance between the atoms $i$ and $j$ and $\theta$ is the angle the transition dipole draws with the vector connecting the two atoms.\\

A crucial property of a laser is its spectrum in the steady state. In order to calculate the spectral distribution of the light field inside the cavity we employ the Wiener-Khinchin theorem~\cite{meystre}, where
\begin{equation}
S(\omega, \, t) =  \int  e^{-i\omega\tau} \, \left \langle a^{\dagger}(t+\tau) a(t)\right \rangle \, \mathrm{d}\tau.
\label{calc_spec}
\end{equation}

Numerically, this is achieved by at first determining the steady state $\rho_S$, which can be calculated as the kernel of the Liouvillian, i.e. solving $\mathcal{L} \left[ \rho_S \right] = 0$. Now, the annihilation operator $a$ is applied and we let this state evolve. After a time $\tau$ has elapsed, we apply the creation operator $a^\dagger$ and Fourier-transform the trace of this aggregate, as the Fourier transformation of the expectation value of the field correlation function equates to the spectrum of the intra cavity and output light field.

\section{Superradiant laser dynamics with confined ensembles}

\subsection{General properties of superradiant lasing}

First, let us exhibit some general features of the dynamics of a laser with all atoms coupled equally to the cavity mode in the two idealized  limiting cases of (a) fully collective and (b) individual independent spontaneous decay. Mathematically, this is implemented simply by setting (a) $\Gamma_{ij} =\Gamma$ for the collective case as discussed in~\cite{haake1993superradiant} and (b)  $\Gamma_{ij} =\Gamma \delta_{ij}$ for independent decay as studied in~\cite{meiser2009prospects}. Surprisingly, the fully collective case is much easier to deal with numerically as the total collective spin magnitude is conserved and the Hilbert space for N atoms is restricted to the $N+1$ states of a  spin-$N/2$ system. The effective pumping of the atoms can also be described as an independent or collective mechanism. Here we refrain from including dipole-dipole induced excitonic shifts of the energy levels. This assumption can be justified for a completely homogeneous atomic density~\cite{intro-haroche} but has to be reconsidered for concrete finite size implementations. We will explicitly account for this in the finite lattice geometries discussed below.

\begin{figure}[h]
\includegraphics[width=4.3cm]{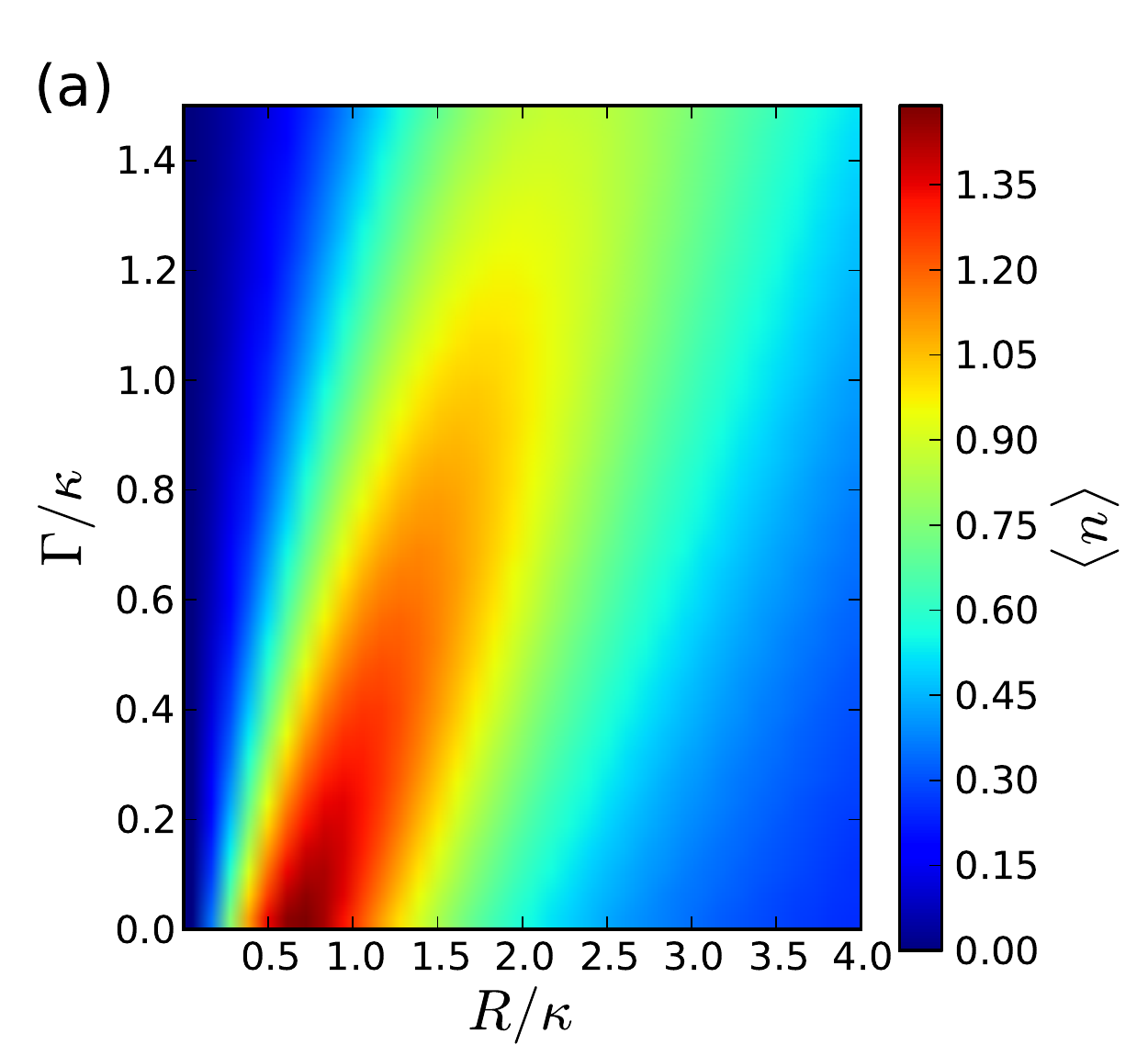}
\includegraphics[width=4.3cm]{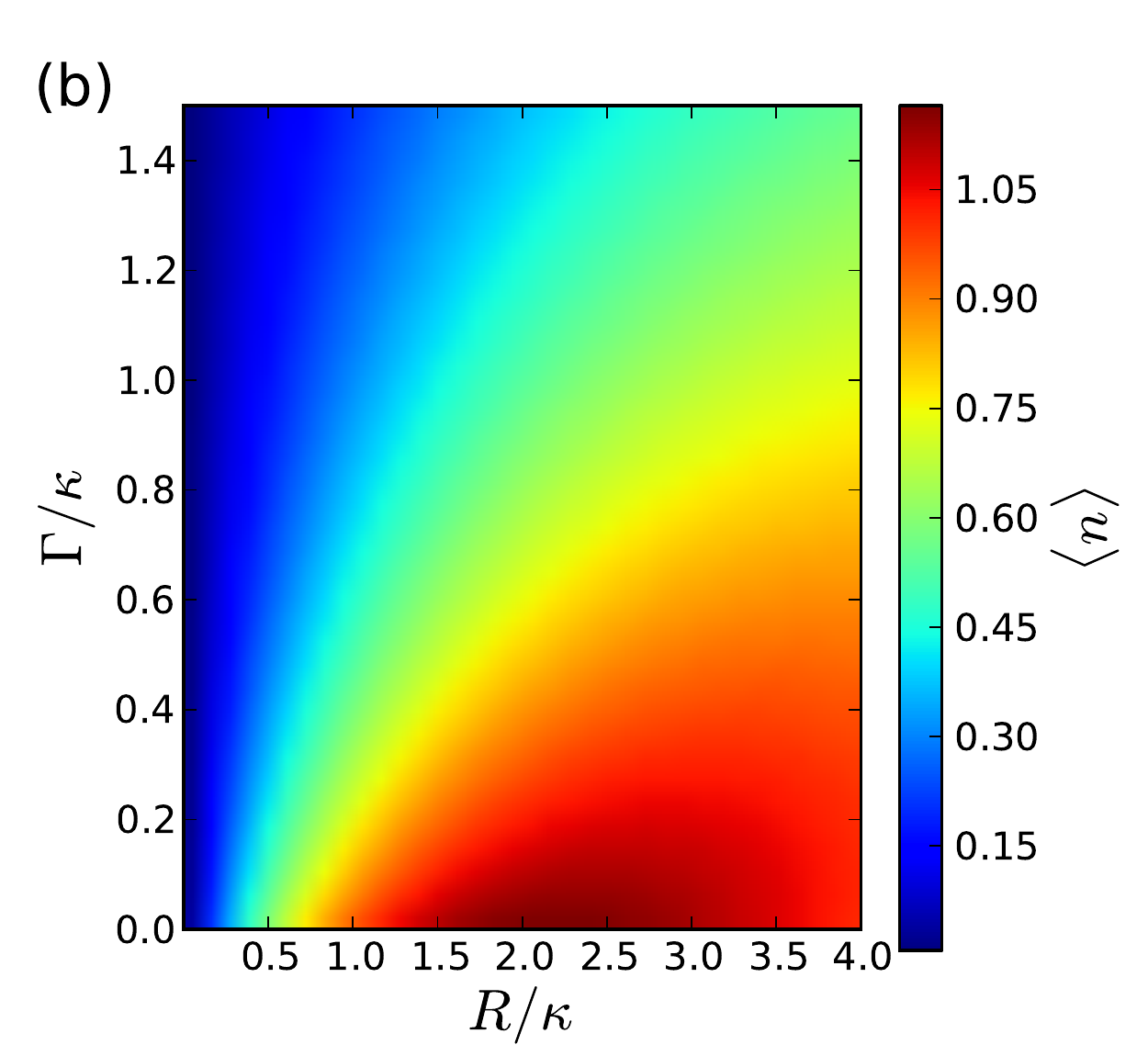}
\includegraphics[width=4.3cm]{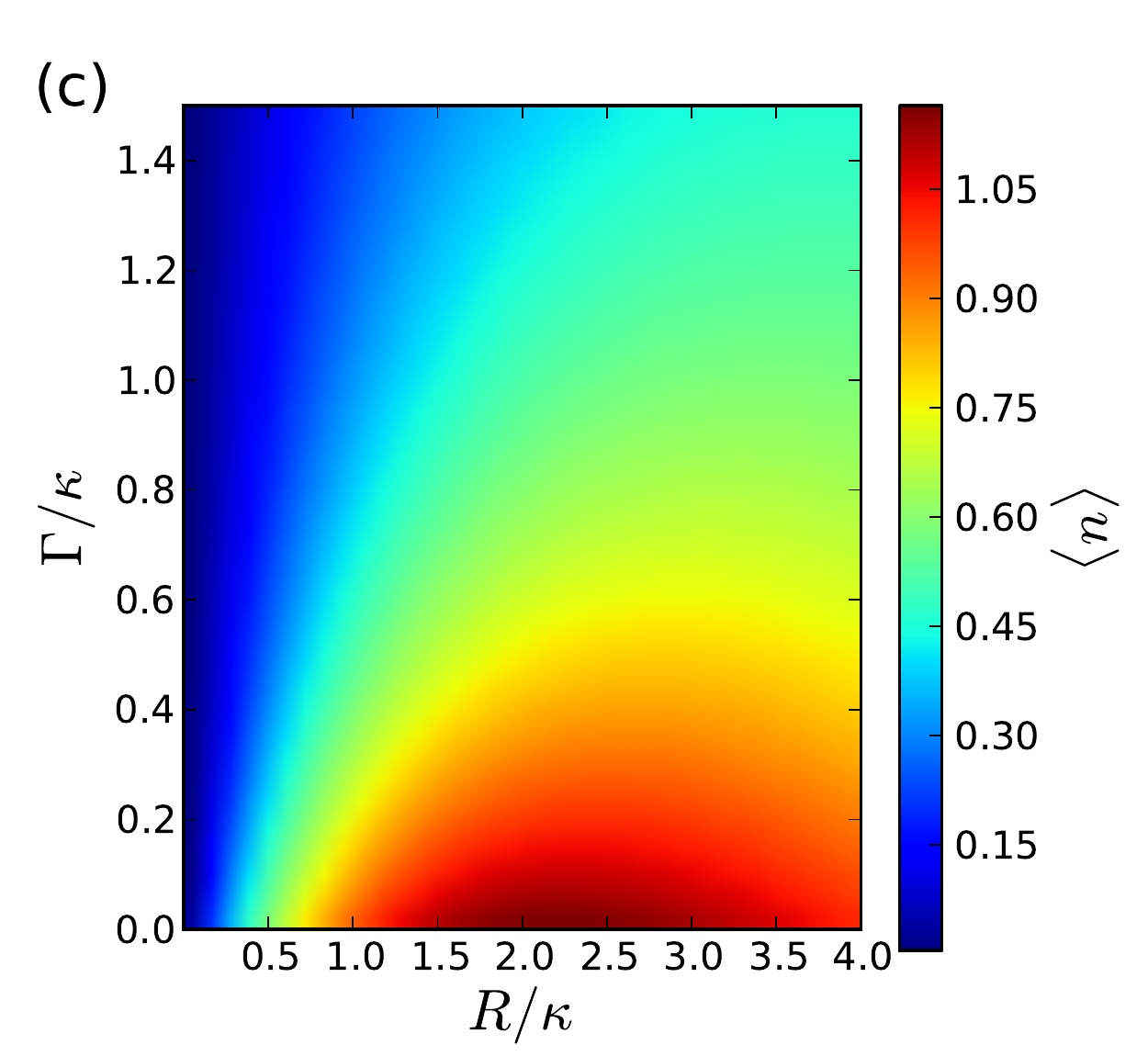}
\caption{Stationary photon number as a function of the pump strength $R$ and the spontaneous decay rate $\Gamma$ for collectively pumped and collectively decaying atoms (a), individually pumped but collectively decaying atoms (b) and individually pumped and individually decaying atoms (c)}
\label{photonnumber}
\end{figure}

Fig. \ref{photonnumber} shows the mean photon number as a function of the pump strength $R$ and the single atom decay rate $\Gamma$ for the three cases of collective pump and collective decay, individual pump and collective decay and independent pump and independent decay for $N=4$ atoms. We see that the maximum photon number is not so different for the three cases and appears at small spontaneous decay rates. For fully collective pump and collective spontaneous decay (fig. \ref{photonnumber}a)superradiant emission into free space limits the optimal operation regime to a lower pump intensity, though.   

\begin{figure}
\center
\includegraphics[width=4.3cm]{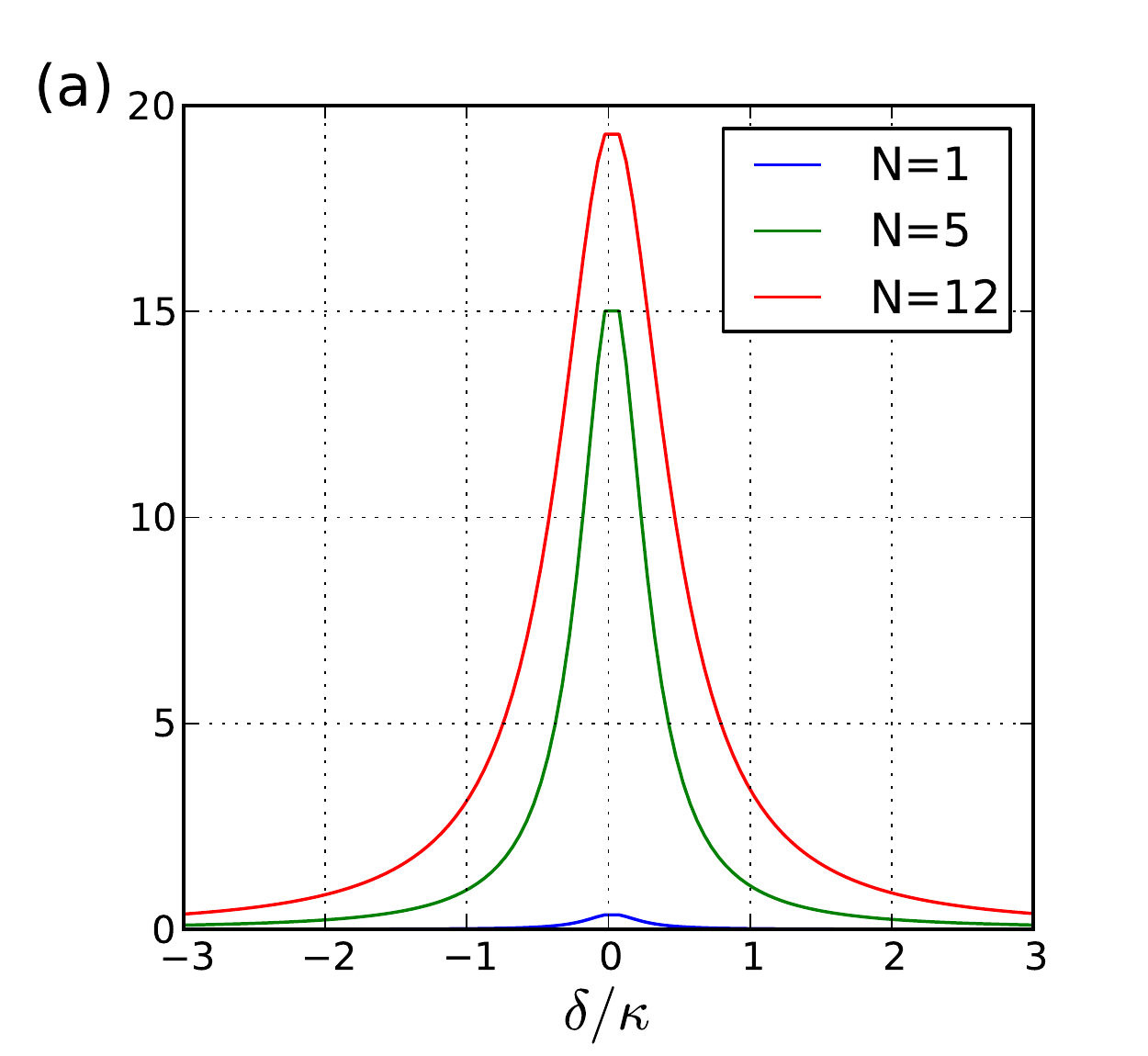}
\includegraphics[width=4.3cm]{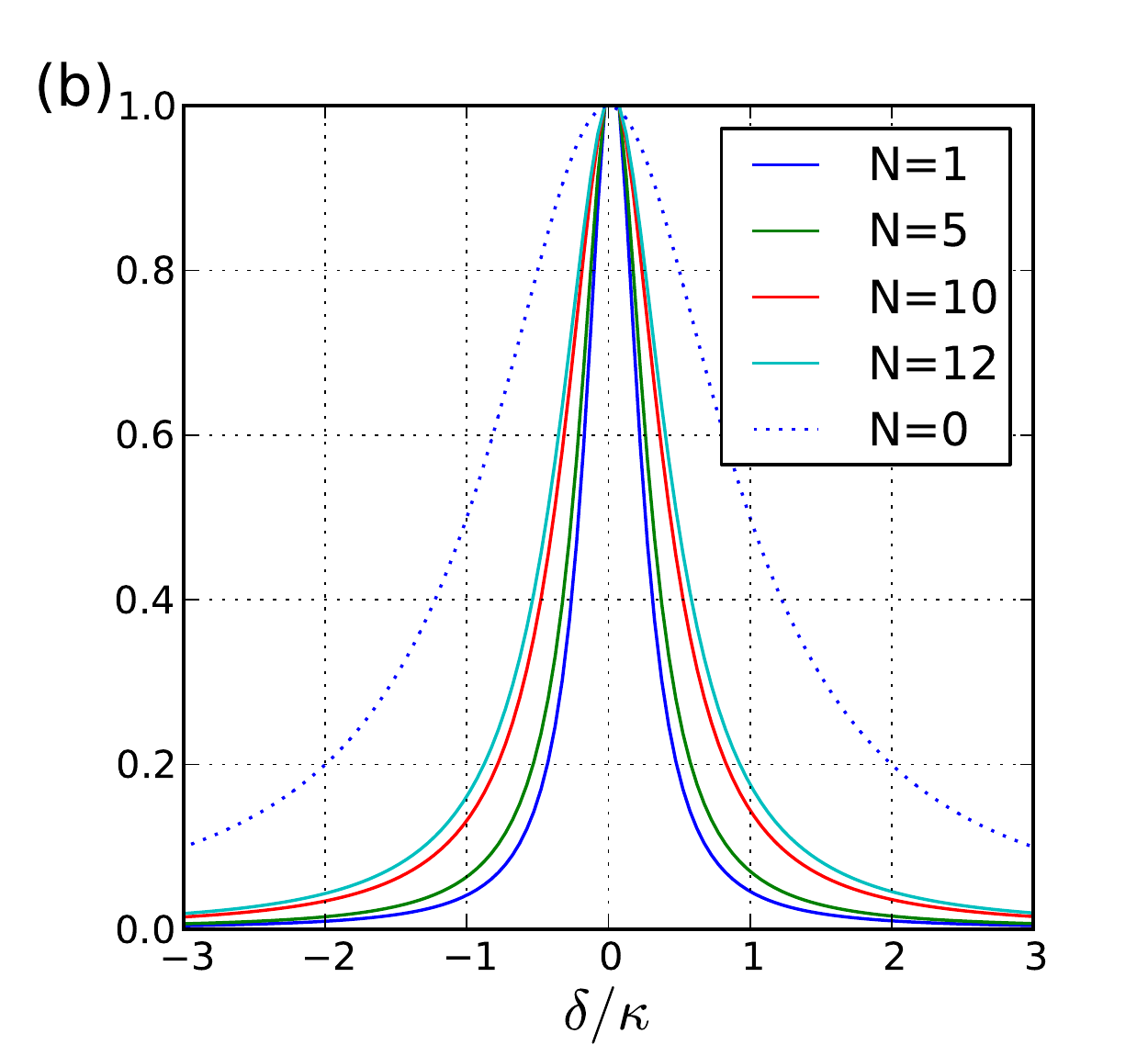}
\caption{Output spectrum of a fully collective laser with different atom numbers $N$ for $\Gamma = \kappa/20$ and $R =  \kappa/5$ compared to the empty cavity linewidth ($N = 0$), absolute (a) and normalized (b)}
\label{spectra}
\end{figure}

Now, it is of course most interesting to look at the frequency stability or line-width of this laser. As seen in fig. \ref{spectra} the output intensity spectrum exhibits a nonlinear growth with the atom number (green line), as expected, until it saturates (red line). Remarkably, however, the linewidth does not narrow with the photon or atom number, but is even increased by superradiant spontaneous emission. Thus, the optimal case seems to be collective emission into the lasing mode without superradiant spontaneous decay. We will investigate this in more detail in the following sections.

\subsection{The superradiant lattice laser}

Above we have seen that collective decay and collective pump strongly change the laser dynamics and its properties. Besides modified decay rates governed by eq.\ref{L_decay} in any finite size geometry dipole-dipole interaction as given by eq.\ref{H_dipole} has to be taken into account as well. To study the basic physical effects, in this section we will investigate three different regular geometric arrangements for the laser active atoms. We compare a linear chain, where we go beyond the single excitation and nearest-neighbor coupling limits discussed in~\cite{zoubi2012collective}, to an equilateral triangle and a square configuration. Let us point out, that for two atoms, e.g.~\cite{intro-dicke}, the particular relative arrangement is irrelevant, and therefore the system can be handled analytically.

\subsubsection{A square lattice of four atoms}

\begin{figure}
\center
\includegraphics[width=4.3cm]{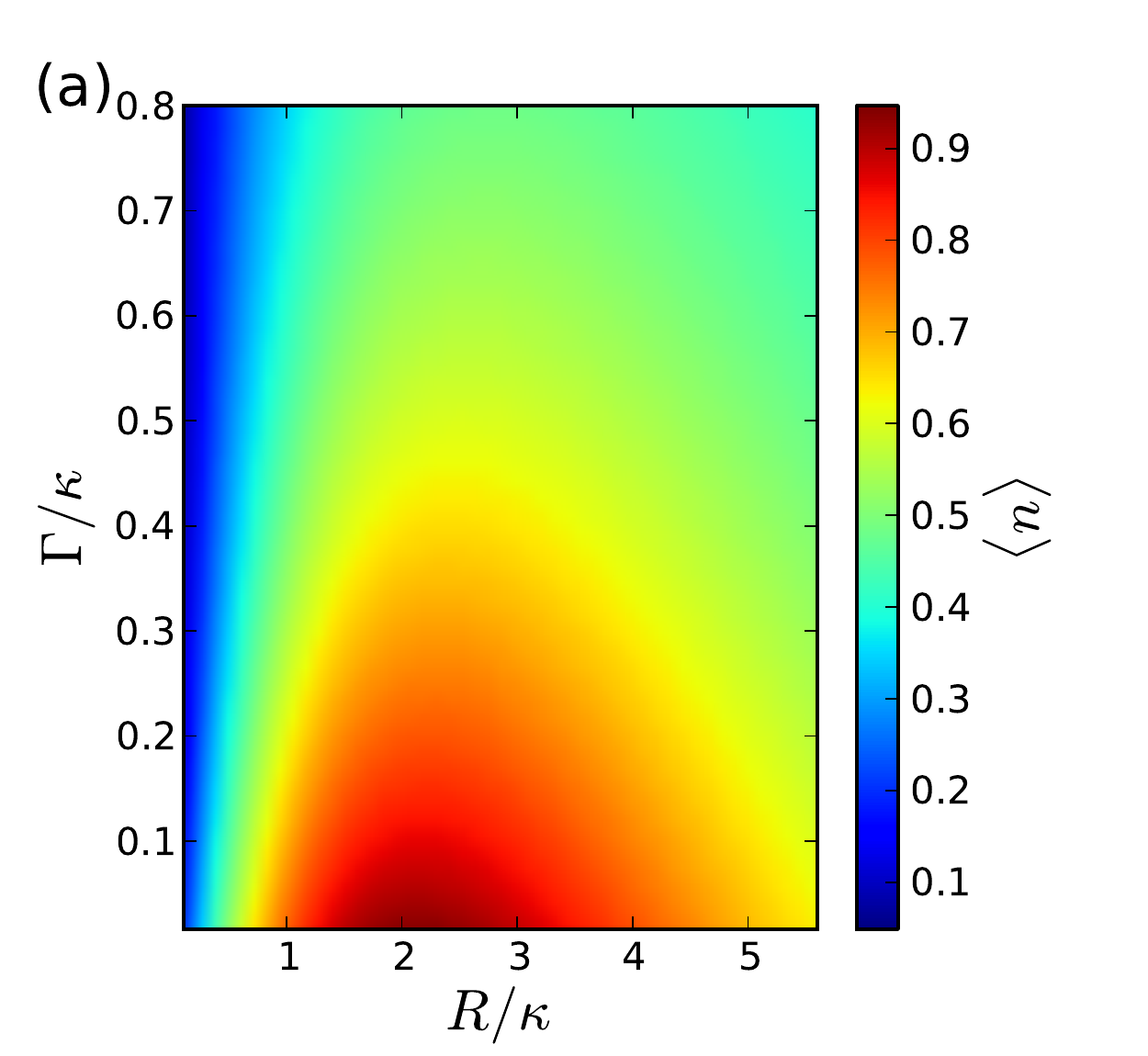}
\includegraphics[width=4.3cm]{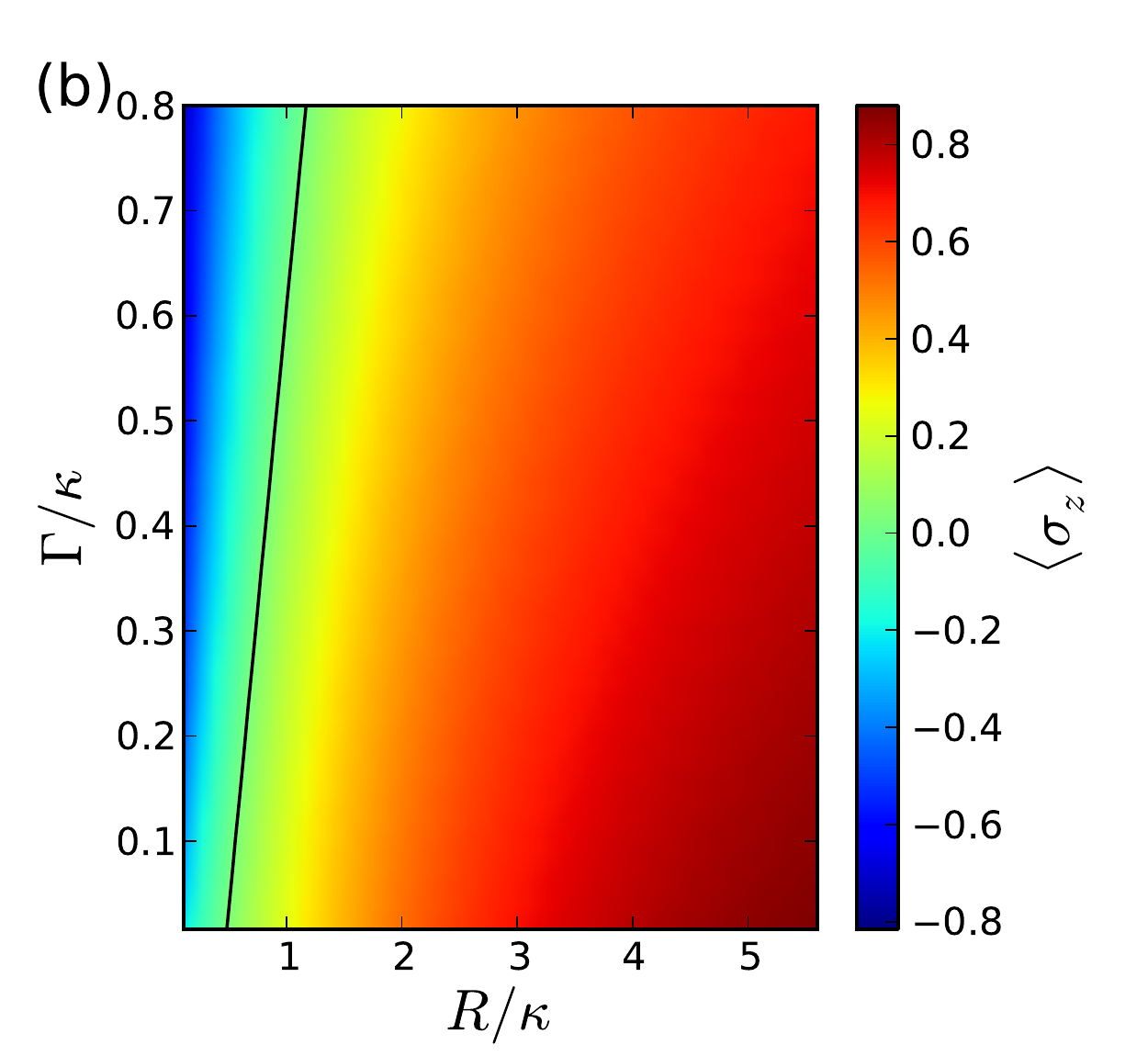}
\includegraphics[width=4.3cm]{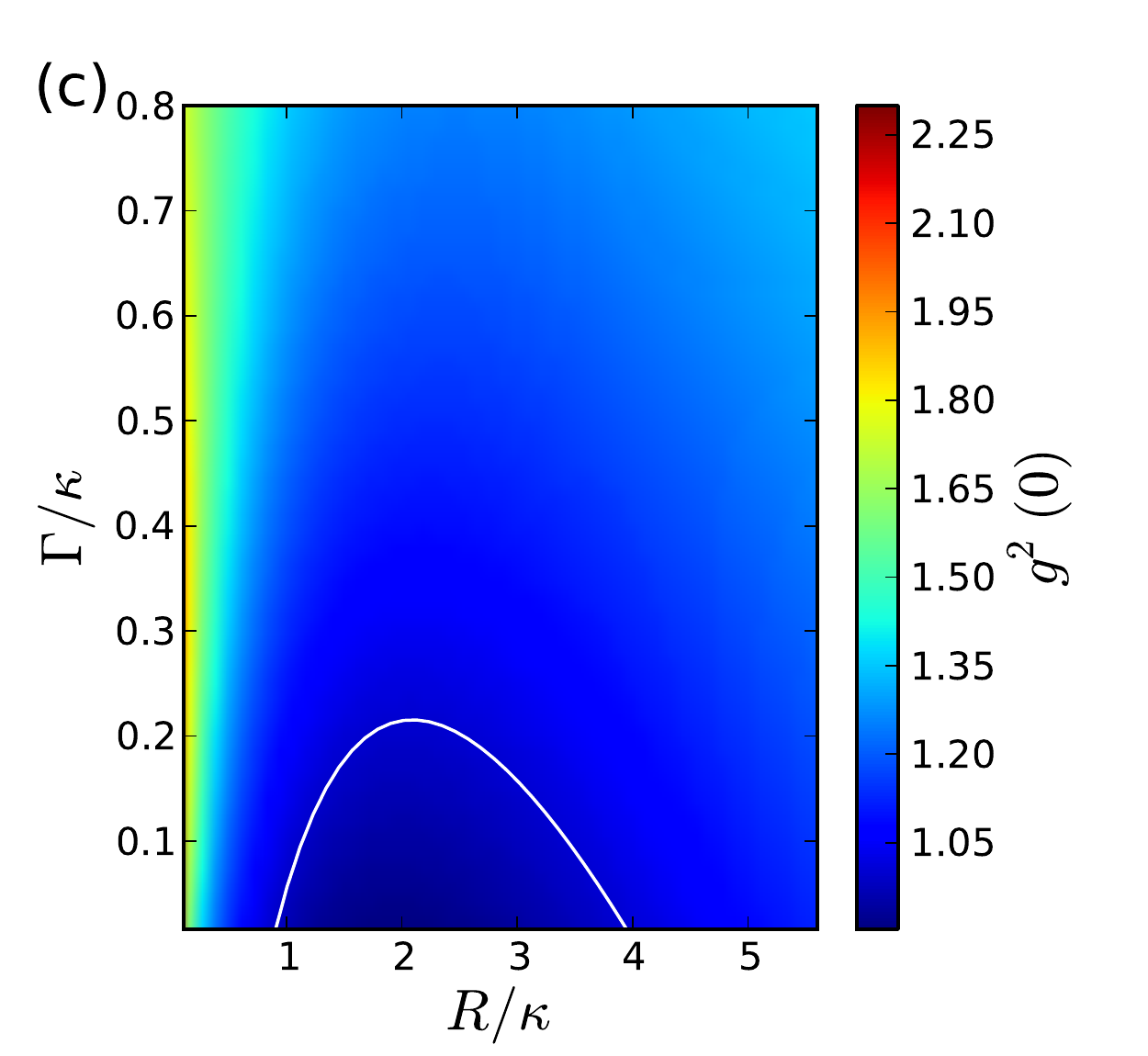}
\caption{Stationary operation of a four atom laser on a square lattice. (a) photon number, (b) atomic inversion, where the black line indicates equal population of the excited and the ground state, (c) $g^2(0)$ function, with the white line at $g^{(2)}(0)=1$  representing a coherent state}
\label{square}
\end{figure}

As a generic example we first show the photon number, the inversion of the active medium atoms and the $g^{(2)}(0)$ correlation function for a fixed cavity loss $\kappa$ while tuning the pumping rate $R$ and the individual atom decay rate $\Gamma$ for a four atom laser in a square lattice. The chosen lattice constant is half of the magic wavelength for Strontium, $\lambda_{magic}/(2 \lambda_0) \approx 0.58$~\cite{Campbell2008, takamoto2005optical}. For the photon number shown in fig. \ref{square}a the maximum appears at a pumping ratio of $R/\kappa = 2.2$, which is equal to the result from above for individual pumping and collective decay as depicted in fig. \ref{photonnumber}b.

In fig. \ref{square}b the expectation value of the $\sigma_z$ operator is illustrated, where the black line represents the crossover to population inversion. On the right-hand side of the line the atom population is inverted, corresponding to the lasing case. Fig. \ref{square}c presents the $g^2(0)$ function, where the white line highlights a value of $g^2(0)=1$, indicating a perfectly coherent light field. The area where $g^2(0) < 1$ could be referred to as an anti-bunching regime.
  
\subsubsection{Comparison of different geometrical configurations}

Let us study the influence of the geometric arrangement of the particles for different numbers of atoms and compare the results for the square discussed above to an equilateral triangle of atoms and a three and four atom chain. In order to obtain a substantial effect despite our small atom numbers, we choose a smaller lattice constant of $a = \lambda_0/10$  and a fixed atomic decay rate of $\Gamma/\kappa = 0.2$.

\begin{figure}
\includegraphics[width=4.3cm]{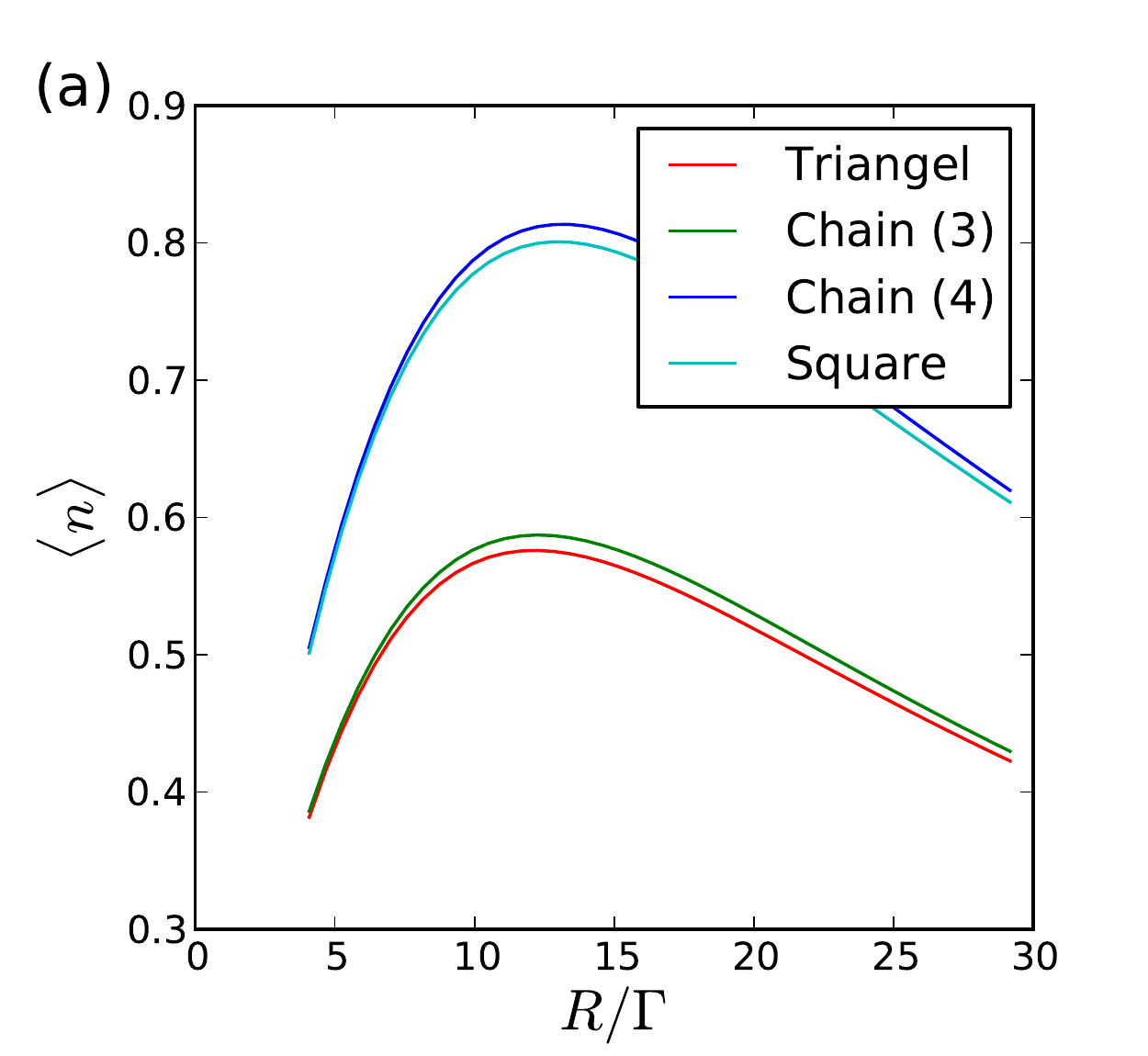}
\includegraphics[width=4.3cm]{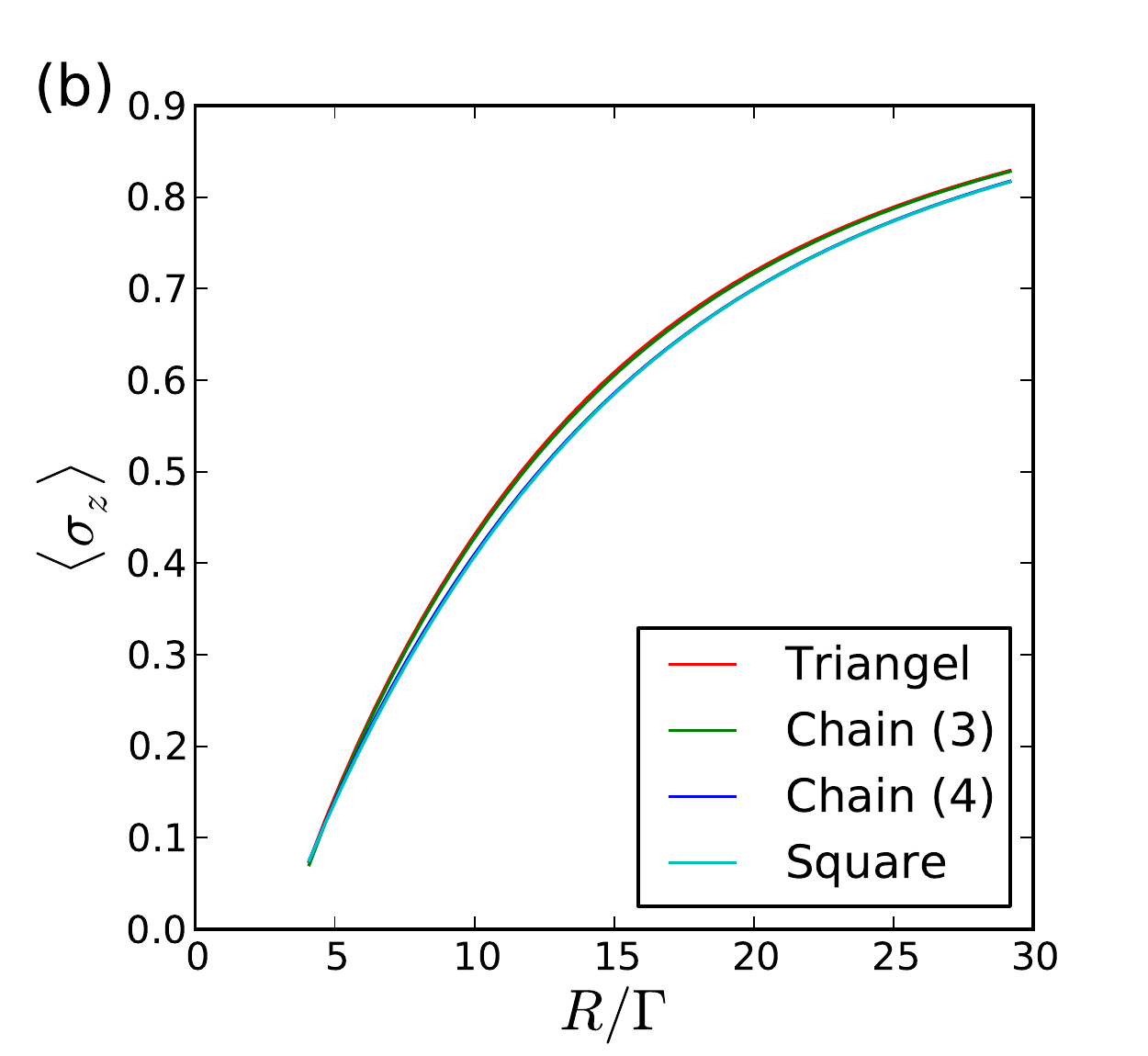}
\includegraphics[width=4.3cm]{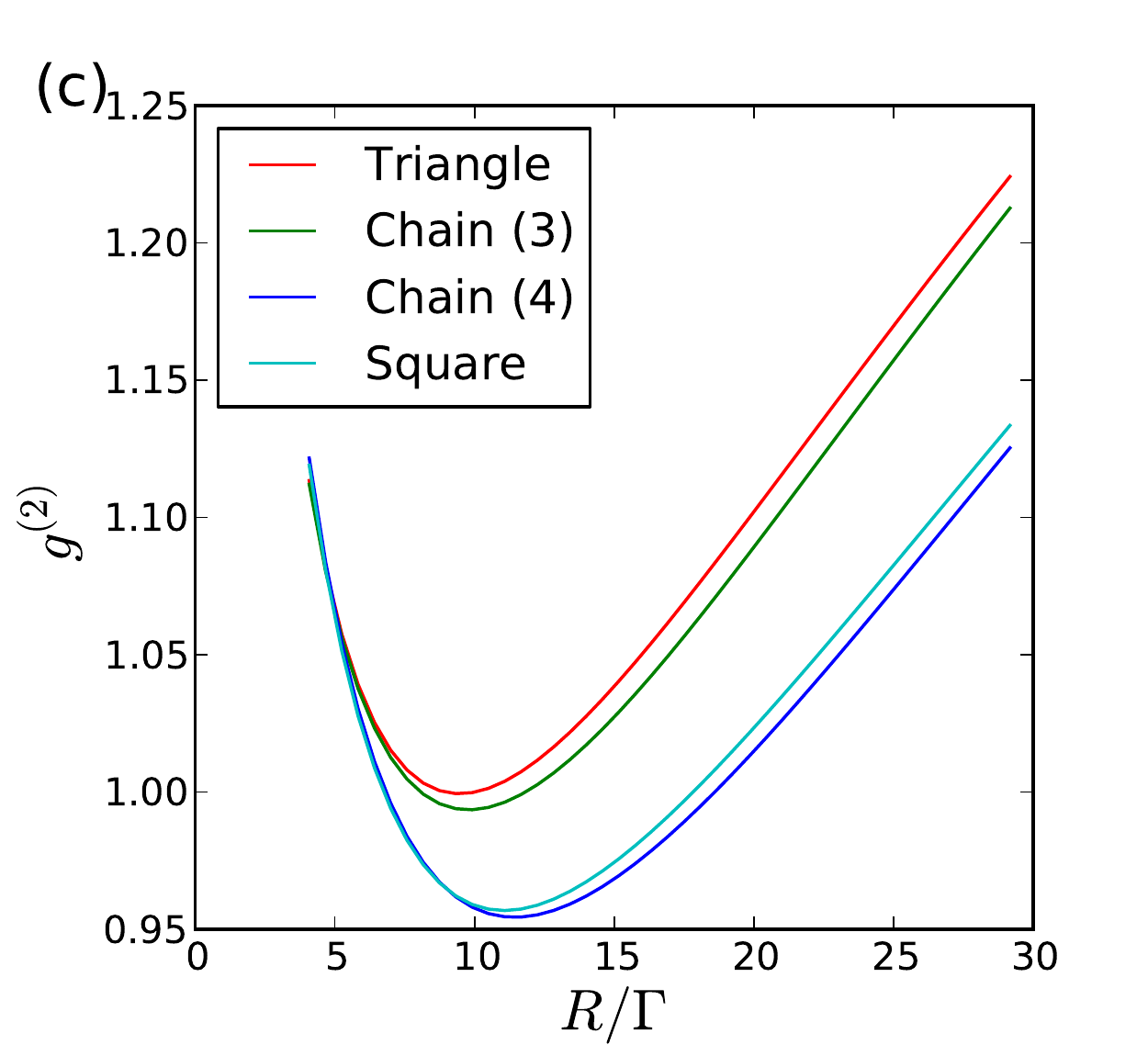}
\caption{Photon number (a), atomic inversion (b) and $g^{2}$ function (c) of the laser as a function of the pump strength $R$ for different atomic arrangements and a fixed spontaneous decay rate $\Gamma=0.2\kappa$}
\label{conf_diff}
\end{figure}

In fig. \ref{conf_diff} we show, that for the average values the atom number is more important than the particular geometric arrangement. Interestingly for four atoms one can even reach sub Poissonian  photon statistics.  

\begin{figure}
\center
\includegraphics[width=4.3cm]{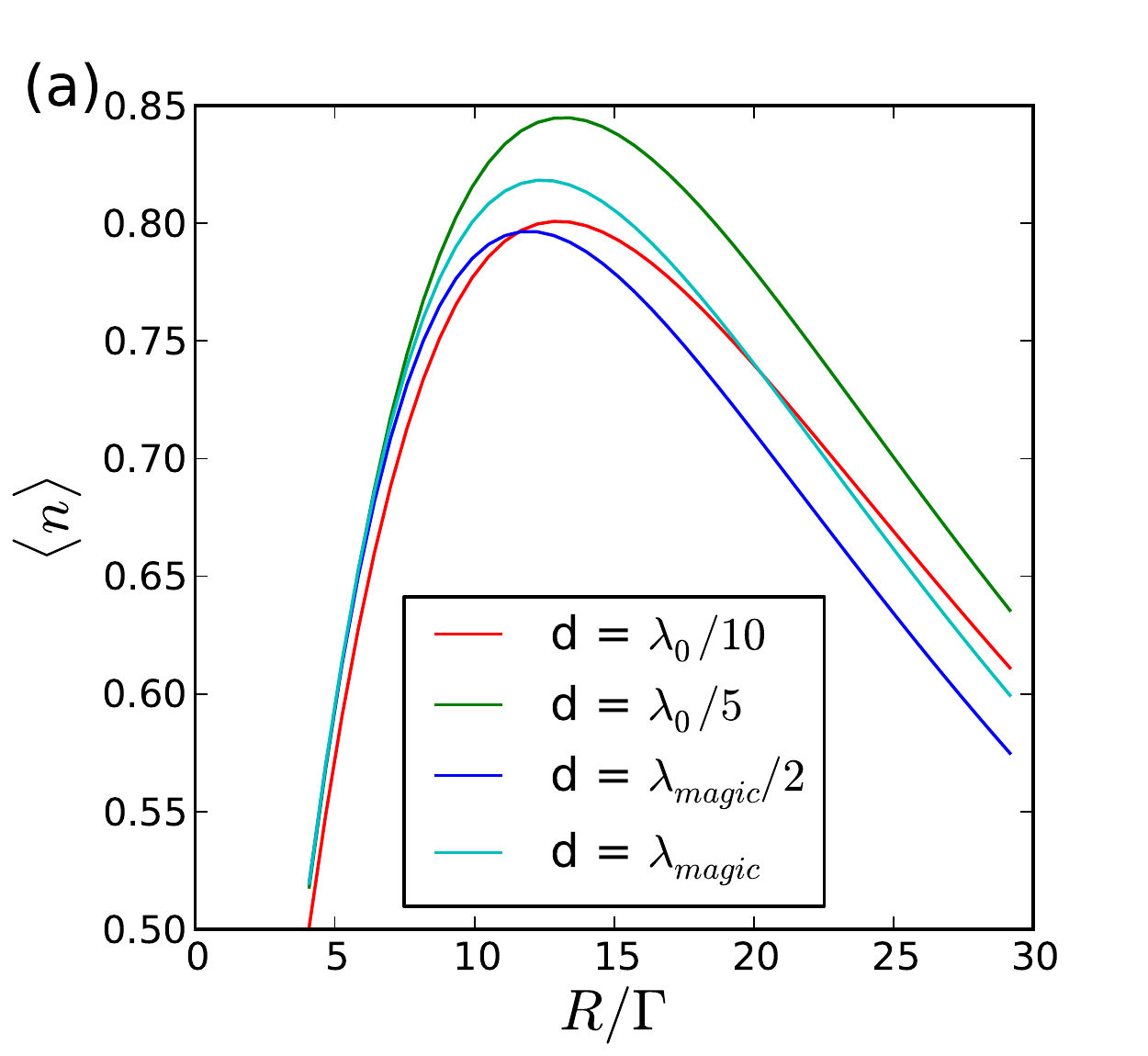}
\includegraphics[width=4.3cm]{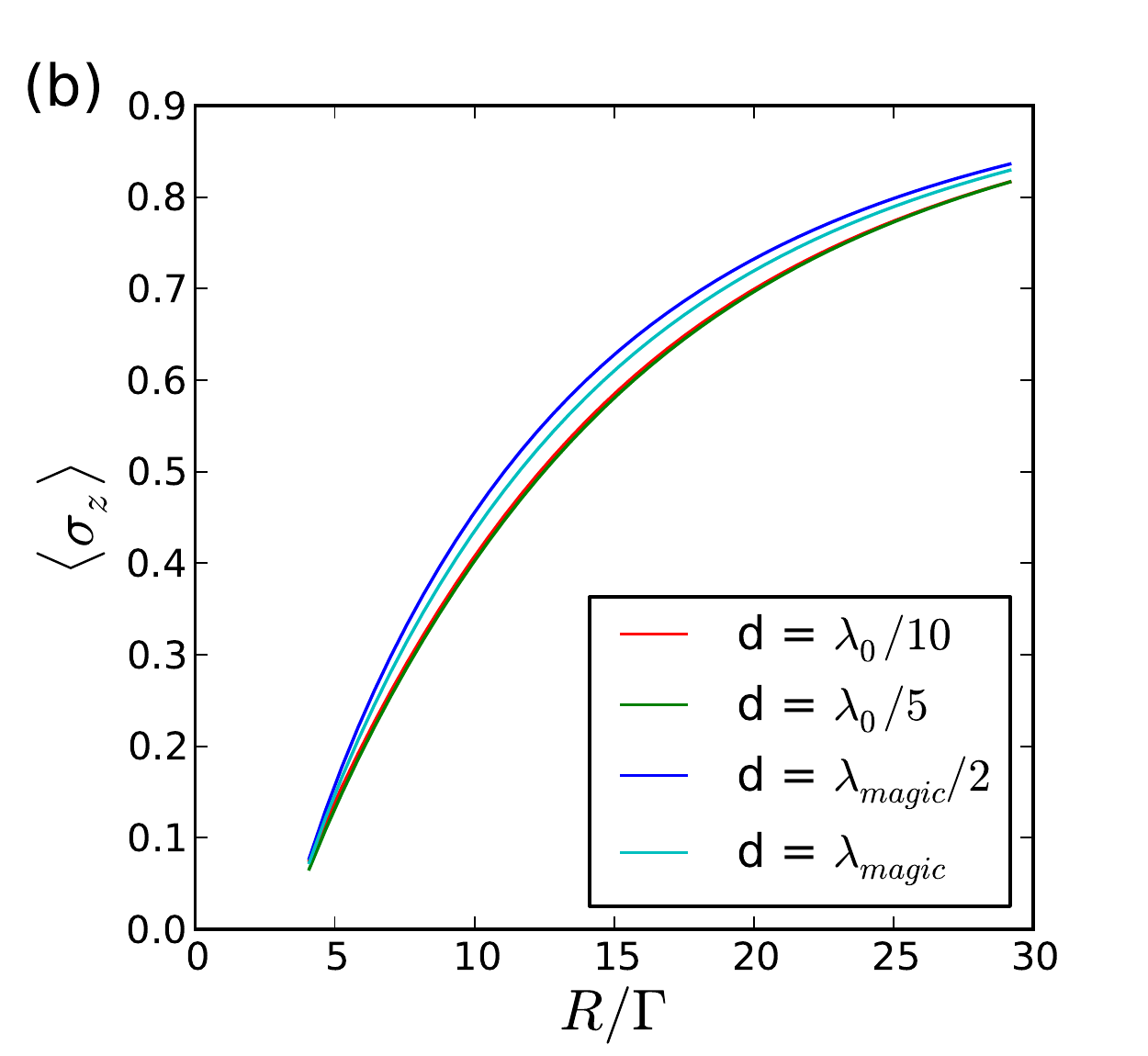}
\includegraphics[width=4.3cm]{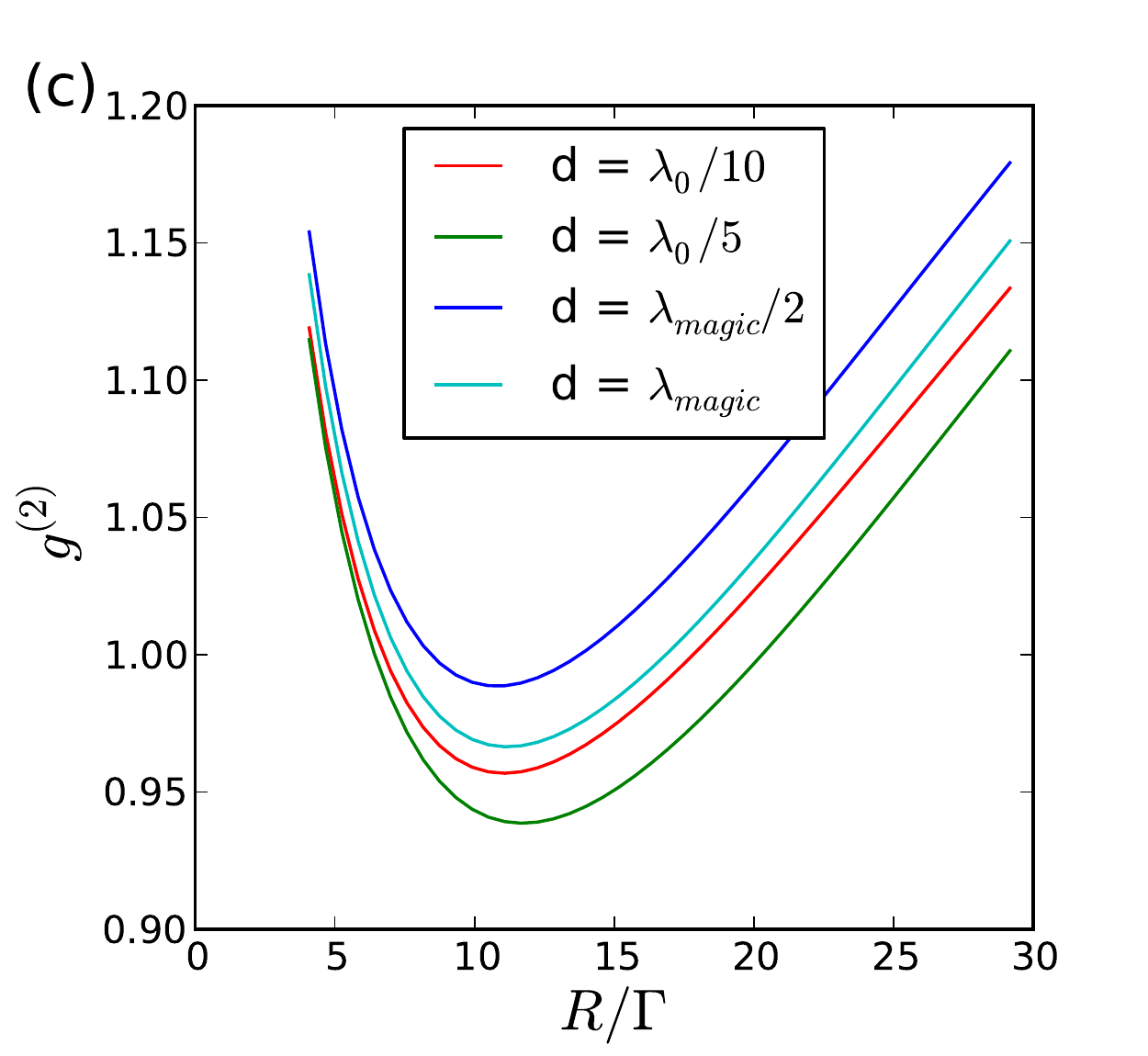}
\caption{Photon number (a), atomic inversion (b) and $g^{2}$ function (c) of the laser as a function of the pump strength $R$ for a square of different lattice constants $a$ and a fixed spontaneous decay rate $\Gamma=0.2\kappa$}
\label{d_diff}
\end{figure}

Naturally, the results depend on the average distance of the atoms, which is shown in the following set of pictures in fig. \ref{d_diff} for a square of different lattice constants $a$ with a fixed spontaneous emission rate of $\Gamma/\kappa = 0.2$. As one might have expected, fig. \ref{d_diff} demonstrates a much more pronounced effect when varying the distance as opposed to changing the geometry.

Overall, despite fairly strong interactions of the atoms at small distances, the laser seems to be very robust against such pairwise perturbations, which appear to average out quite well once  the oscillation threshold is surpassed. The differences increase with pump strength where, on average, more particles are excited.

\section{Laser stability and frequency shifts for different atomic distances}

Of course, the most sought after quality of a superradiant laser is its superb frequency stability and accuracy. In the first section we have seen that collective spontaneous decay can broaden the laser line and dipole-dipole interactions potentially shift its position. Now, we will study this effect for a lattice laser.

\subsection{Laser linewidth and frequency shift}

\begin{figure}
\center
\includegraphics[width=5cm]{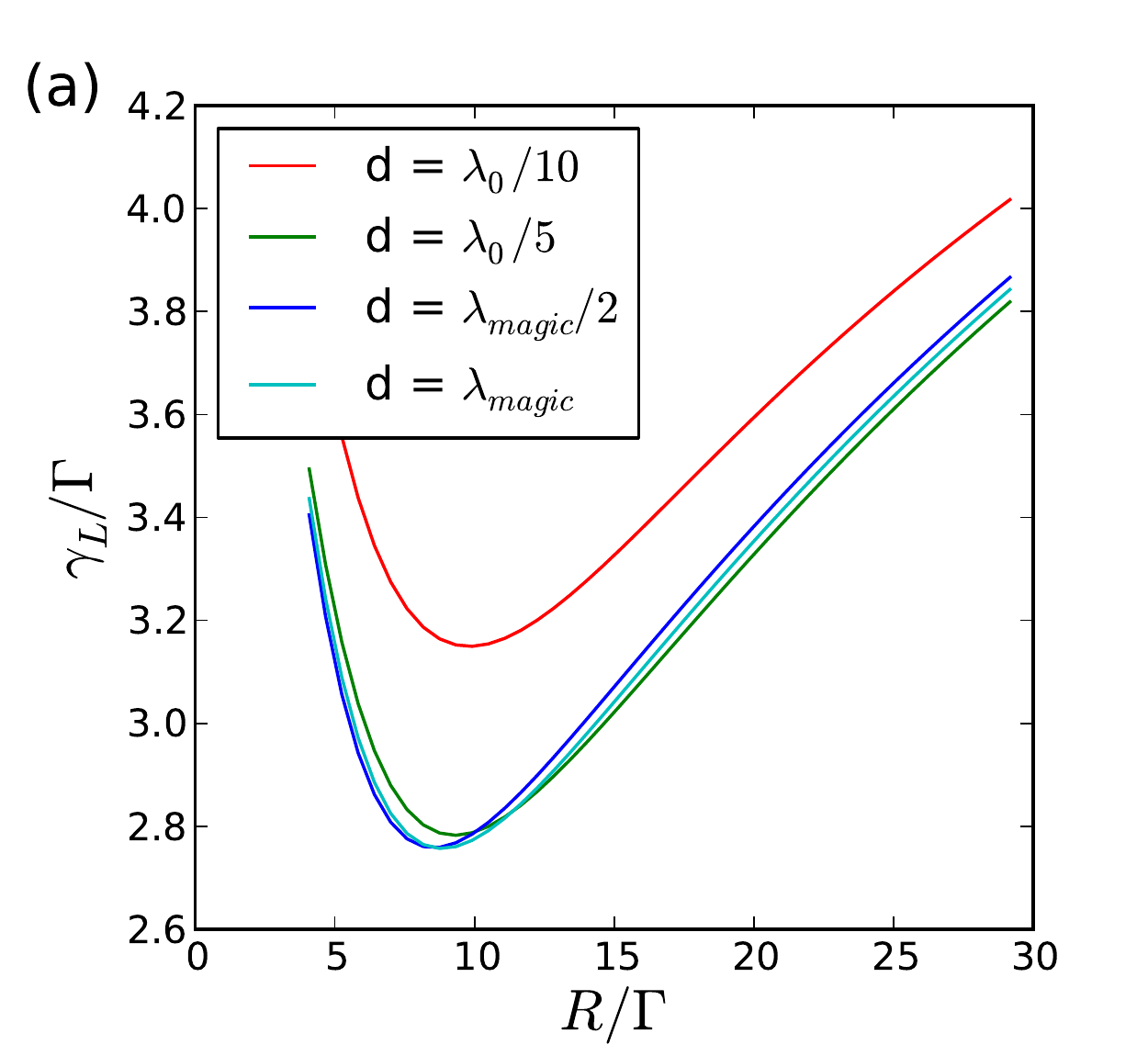}
\includegraphics[width=5cm]{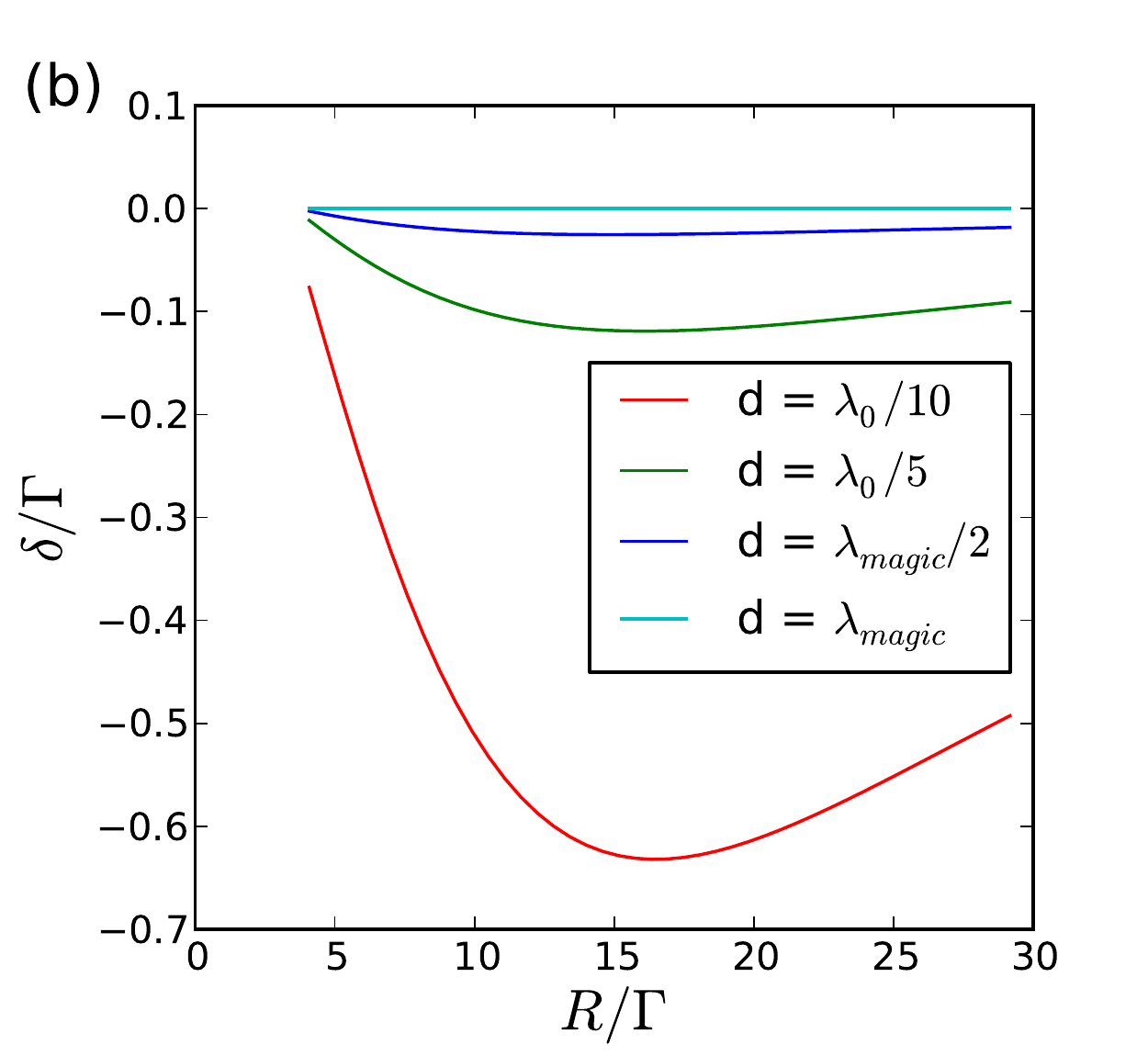}
\caption{Laser linewidth (a) and frequency shift (b) for a square atom arrangement at different distances as a function of the pump strength for a fixed atomic decay rate of $\Gamma=0.2 \kappa$}
\label{lwsquare}
\end{figure}

\begin{figure}
\center
\includegraphics[width=5cm]{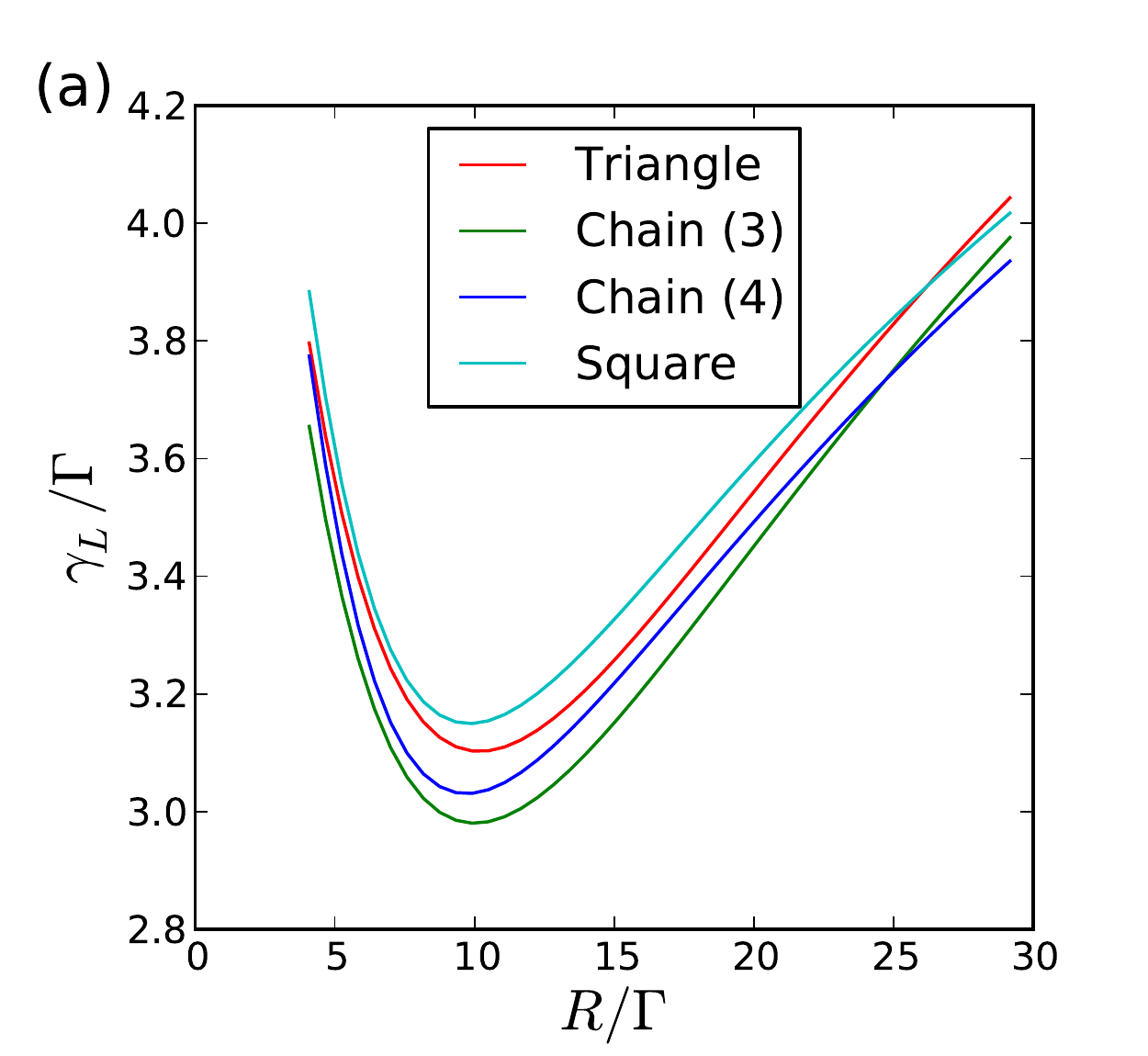}
\includegraphics[width=5cm]{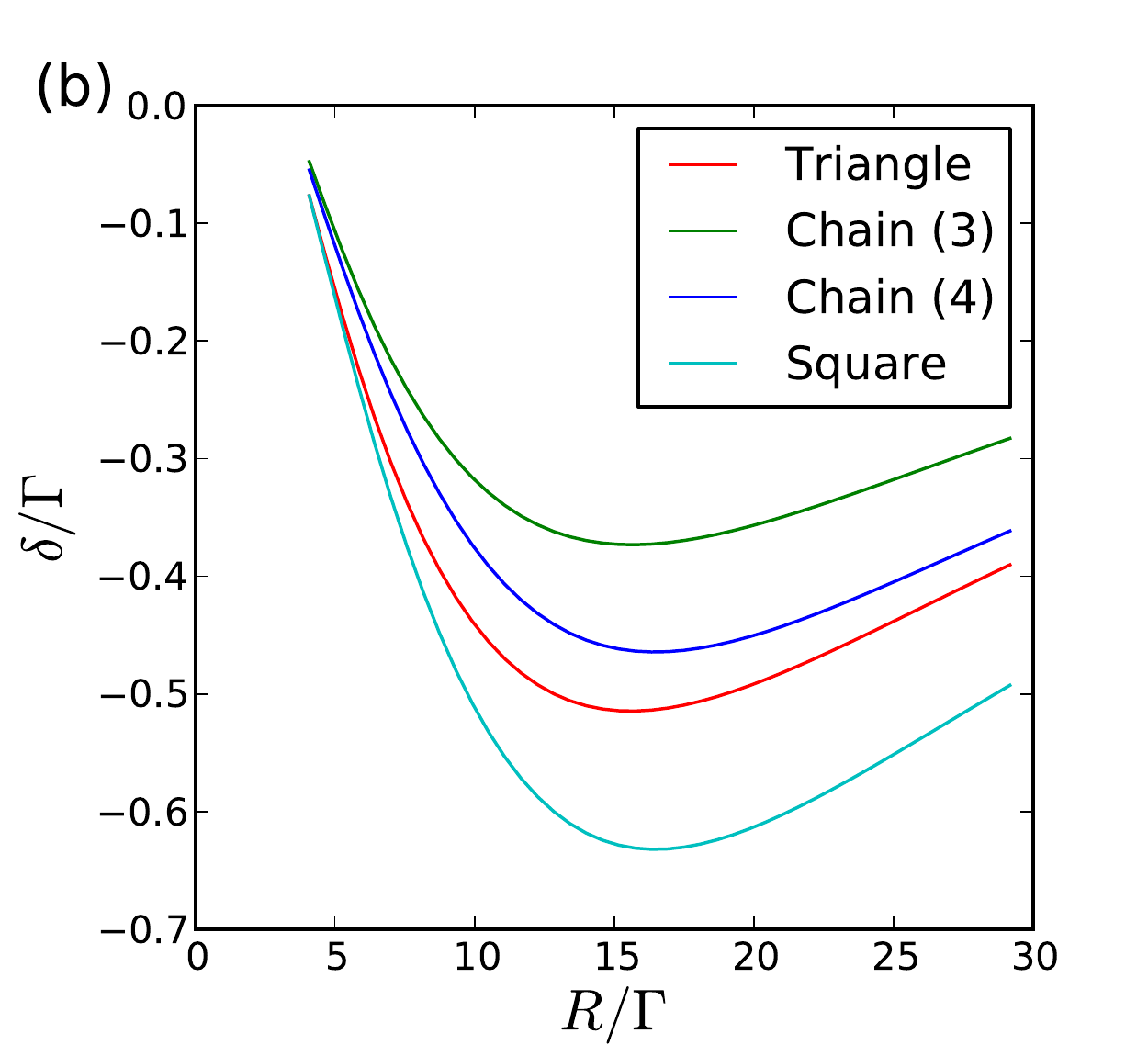}
\caption{Laser linewidth (a) and frequency shift (b) for different geometric configurations as a function of the pump strength for a fixed atomic decay rate of $\Gamma=0.2 \kappa$}
\label{lwgeo}
\end{figure}

 As is well known, the spectrum of a laser in the bad cavity limit deviates from the idealized Shawlow-Townes result, but the center of the line still approximately follows a Lorentzian~\cite{henschel2010cavity} so that in our numerical analysis the linewidth and its center position relative to the bare atom line can be determined from a Lorentzian fit to the steady state spectrum, as described in sec. 2. Therefore, the width of the Lorentzian corresponds to the laser's linewidth while the offset in the maximum describes the energy shift, which is the energy of the light field in the cavity relative to the cavity ground frequency. Fig. \ref{lwsquare} and  fig. \ref{lwgeo} present the fitted width $\gamma_L$ and the energy shift $\delta$ for different interatomic distances and geometrical configurations as a function of the pumping rate $R$. For these calculations the same parameters as above were used.

In fig. \ref{lwsquare} we depict the linewidth and frequency shift of a laser with four atoms in a square configuration as a function of the pump strength for different inter atomic distances.   We observe a minimum linewidth at a moderate pump strength of $R/\kappa \approx 1.9$, which corresponds to an operation at the maximally achievable photon number, as shown in fig. \ref{square}.

For a stronger pump the perturbations due to collective interactions dominate, though significant effects appear for very closely positioned atoms, i.e. $a<\lambda/2$, only. Even with just four atoms it is possible to achieve a linewidth significantly below the resonator's linewidth. The predicted frequency shift with respect to the bare atom frequency (as depicted in fig. \ref{lwsquare}) remains very small for larger interatomic distances and reaches a maximum value when  the laser is operated at  $R/\kappa \approx 3$, close to the maximum photon number.  This could certainly be an observable phenomenon, but it is not detrimental for the operation of such a laser.

Interestingly, for the linewidth and shift properties, geometrical effects are more important than they are for the average intensity. A square arrangement of the atoms creates a much larger shift than a triangular or a linear array, as can be seen in fig. \ref{lwgeo}. Note that the increased shift with the atom number could lead to observable perturbations for larger ensembles. Again, operation at a lower pump intensity could help to minimize the effect.

\subsection{Laser sensitivity to cavity length fluctuations}

\begin{figure}
\center
\includegraphics[width=5cm]{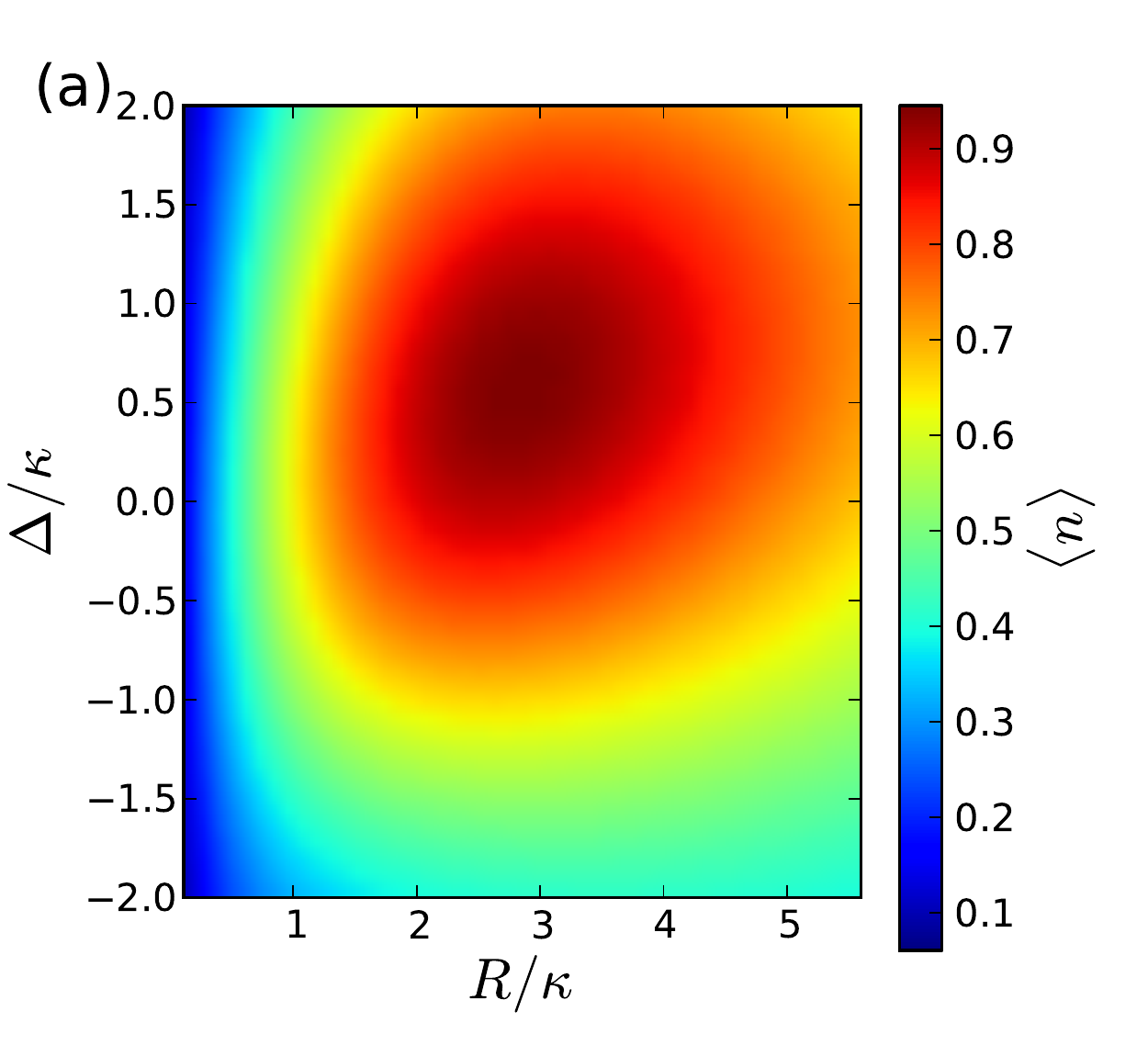}
\includegraphics[width=5cm]{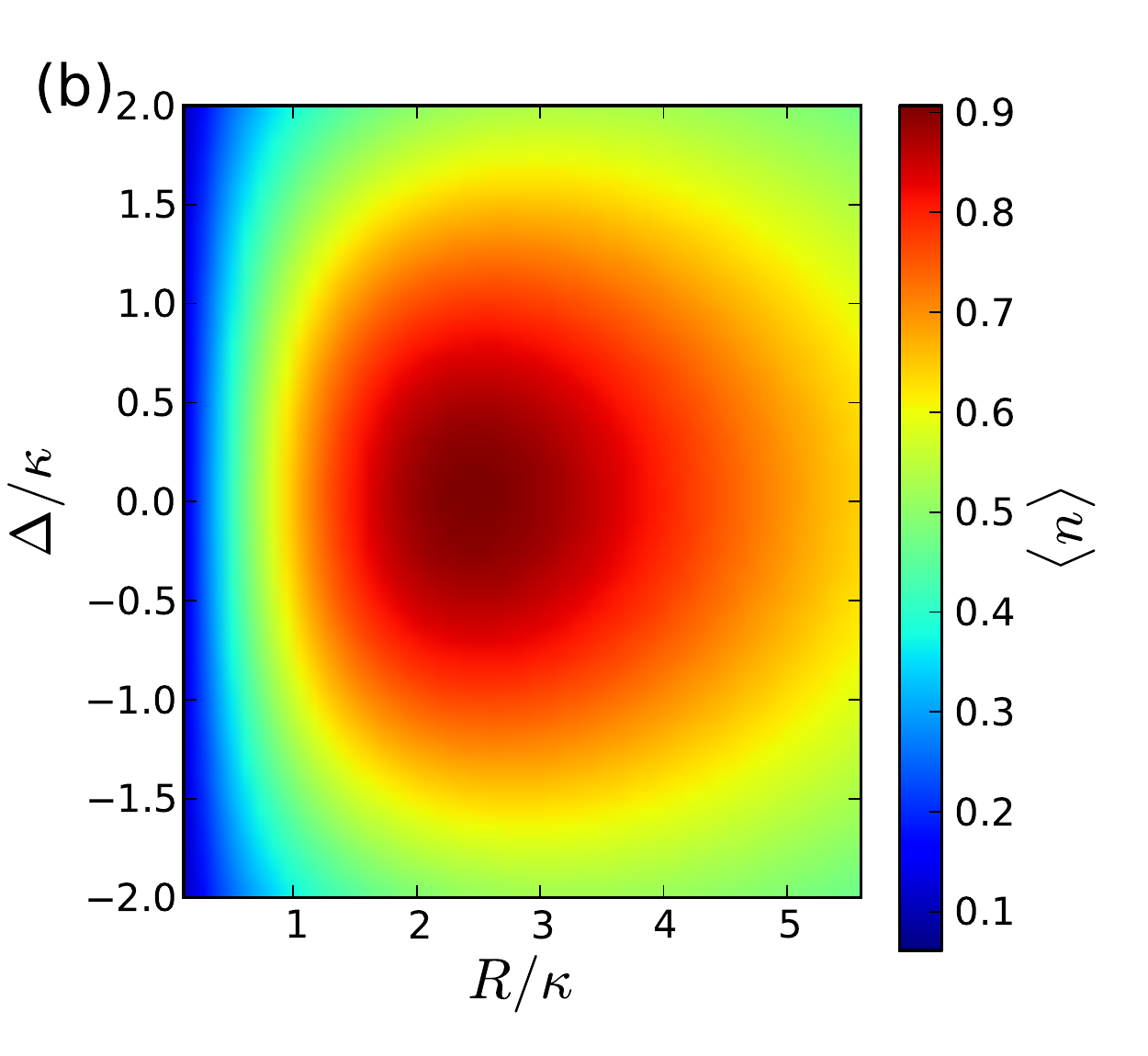}
\caption{Average photon number for atoms on a square with $a = \lambda_0/10$ (a) and $a = \lambda_{magic}/2$ (b)  for variable cavity detuning and an atomic decay rate $\Gamma=0.2 \kappa$}
\label{numdelta}
\end{figure}

\begin{figure}
\center
\includegraphics[width=5cm]{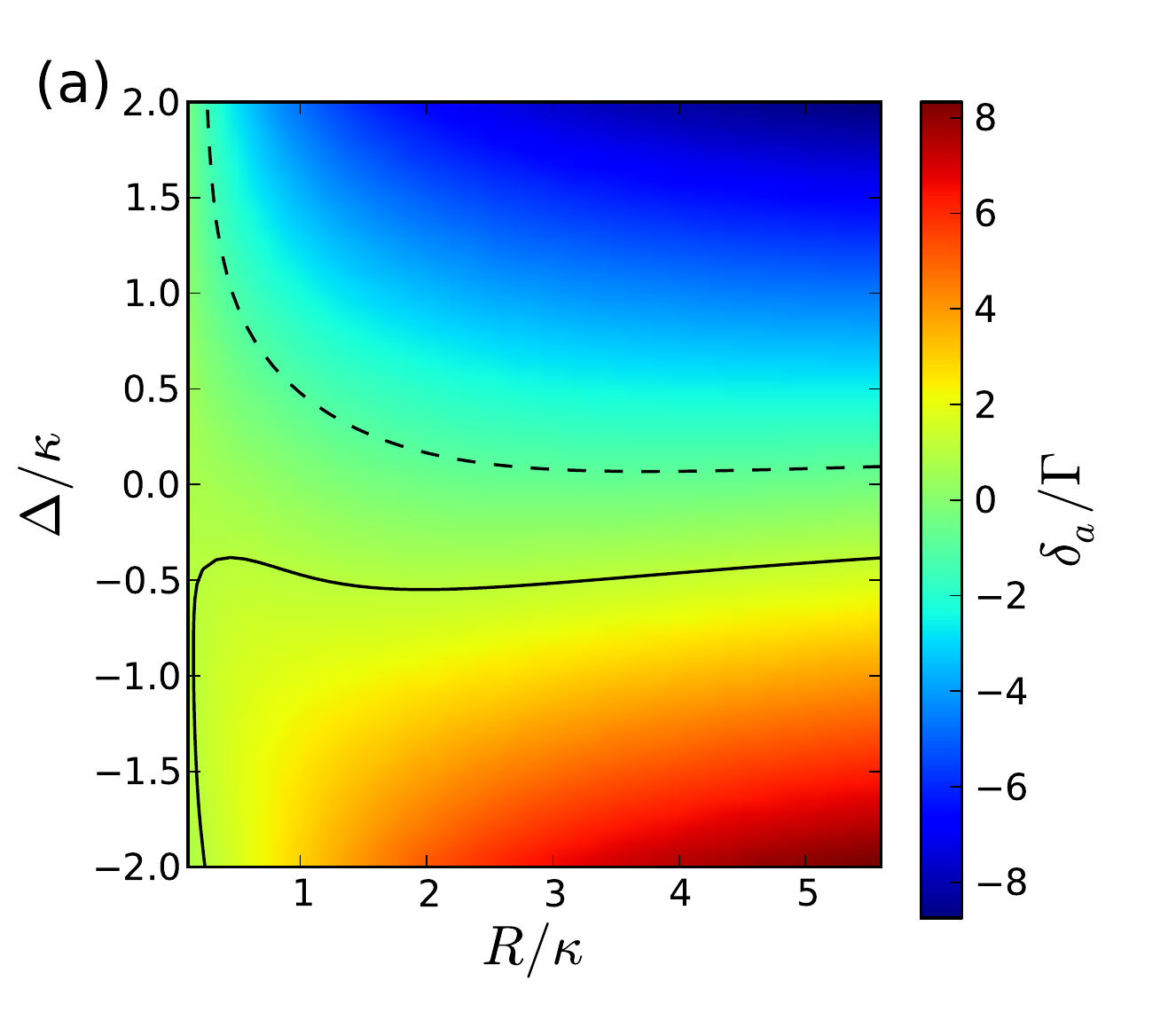}
\includegraphics[width=5cm]{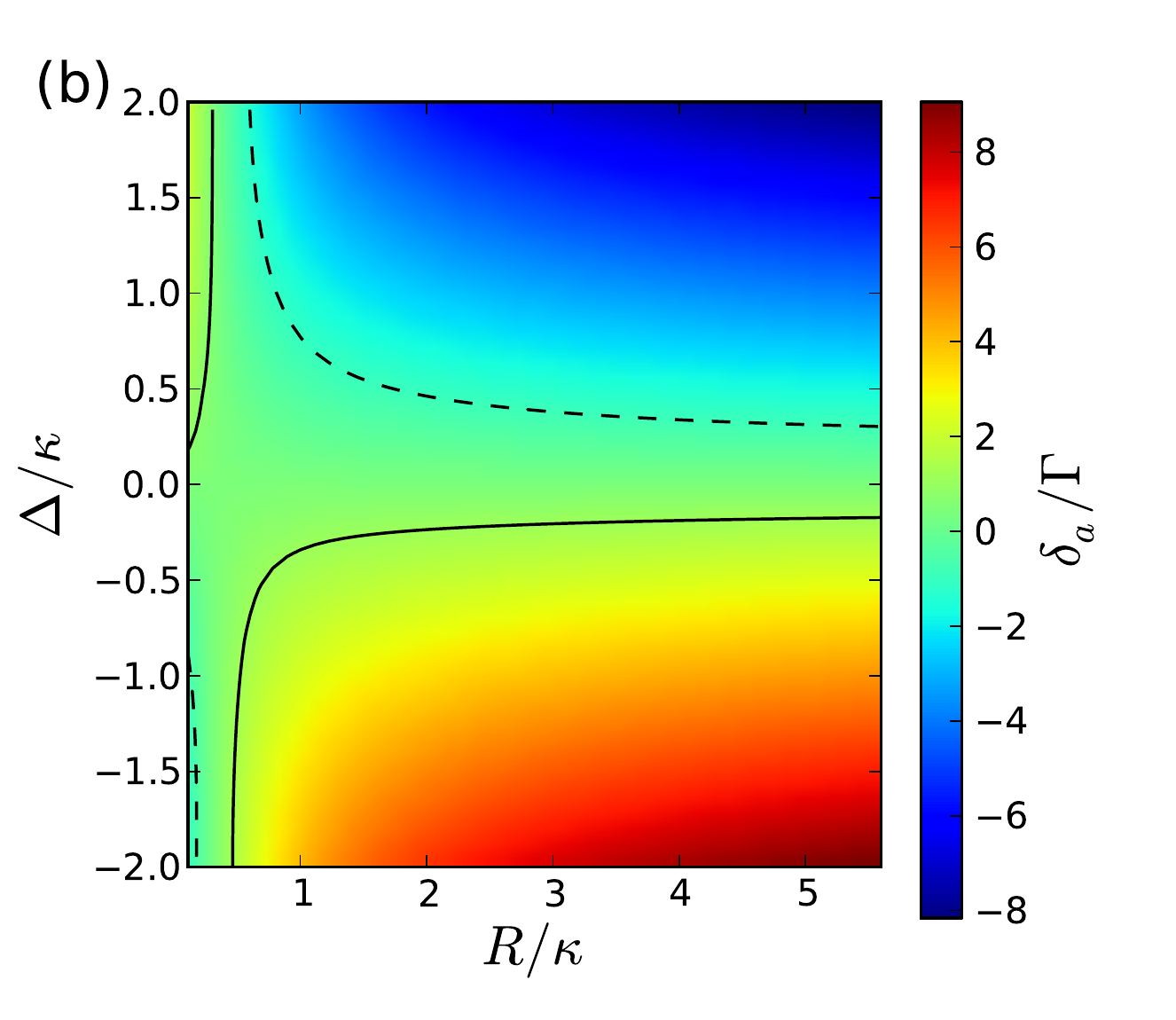}
\caption{Frequency shift for a square atom configuration with $a = \lambda_0/10$ (a) and $ a = \lambda_{magic}/2$ (b)  for variable detuning and a fixed atomic decay rate of $\Gamma=0.2 \kappa$. The dashed line represents $\delta_a/\Gamma = -1$ and the solid line corresponds to $\delta_a/\Gamma = 1$}
\label{Xdelta}
\end{figure}

A central criterion for the stability of a laser is its sensitivity to fluctuations of the effective cavity length, which at present is one of the main limitations of reference oscillator stabilized lasers. Despite spectacular recent progress~\cite{bloom2014optical}, comprehensive control at this level  is still an extraordinary technical challenge. With the atoms acting as reference oscillators less effort in order to achieve technical stabilization is expected in an ideal superradiant laser. In the following we will study the effect of a varying cavity frequency described by an effective detuning ($\Delta = \omega_c - \omega_0$) on the average photon number (fig. \ref{numdelta}) and the frequency mismatch between the bare atomic transition frequency and the laser field ($\delta_a = \omega_0 - \omega_L$) as seen in fig. \ref{Xdelta} depending on the average atomic distance. As shown in fig. \ref{numdelta}a for closely positioned atoms the interaction evokes a significant blue shift of the cavity frequency, generating the maximum photon number with respect to the clock transition. For atoms in a magic wavelength lattice (fig. \ref{numdelta}b) this shift is much smaller and close to the interaction-free case. The detuning sensitivity of the laser output spectrum in these two cases is depicted in fig. \ref{Xdelta}. 

We see that the laser frequency pulling via the cavity changes with the interaction and increases with pumping and the intracavity photon number. Nevertheless, as indicated by the solid and dashed lines, the effective laser frequency change remains within an atomic linewidth even for cavity fluctuations on the order of the cavity width. At low pump strength and small inversion a sort of self-synchronization of the atomic dipoles via direct interactions can lead to very strong suppression of cavity fluctuations at the expense of very little output light, while for stronger pumping interaction effects are suppressed and the cavity drifts produce a more significant impact on the laser frequency. Overall, we observe that by choosing optimal operating conditions a decoupling of the cavity fluctuations from the laser frequency can be suppressed very effectively, even in the case of atomic interactions. However, this decoupling generally also reduces the output power of the laser.

\section{Conclusions and outlook}

By means of numerically solvable examples involving a few particles only, we have evaluated the influence of dipole-dipole interaction and collective spontaneous emission on the radiative properties of a superradiant laser in a lattice geometry. In general, even for fairly closely spaced atoms, shifts and frequency uncertainties are of the order of the free space atomic linewidth. Only for very densely packed ensembles superradiant free space decay will substantially broaden the laser line and increase the sensitivity of the  laser frequency to cavity drifts. Quantitatively, the various limiting cases of a completely collective laser as opposed to an independent atom system can lead to a different scaling behavior of the photon number and the linewidth with substantially different photon statistics. Fortunately, for a Strontium setup based on a magic wavelength lattice, the detrimental effects remain very small, although they could attain an observable magnitude at high filling. At optimally chosen operating conditions dipole dipole interaction can be exploited to reduce laser frequency fluctuations via direct phase stabilization even at very low photon numbers.
 
In this work we still assumed a rather ideal and to some extent artificial pumping mechanism, replenishing the upper atomic state by introducing the minimum necessary decoherence only, while also neglecting light shifts from the pump lasers. Any more realistic pumping via extra levels or an injection of excited atoms would, of course, add extra noise and has to be designed very carefully. This is one of the major remaining challenges for the implementation of such an optical version of the hydrogen maser. In any case, from the point of view of stability and shifts, the operation at weak pump strength seems favorable, although the very weak output field could be a technical challenge for practical use.          

\section*{Acknowledgements}

We thank M. Holland, D. Meiser, J. Ye, C. Genes  and A. M. Rey for stimulating discussions and acknowledge support by DARPA through the QUASAR project and the Austrian Science Fund via the SFB Foqus project F4013.

\end{document}